%% file: main.tex
\documentclass[twocolumn,twocolappendix,usenames,dvipsnames]{aastex63}

\usepackage{amsmath}
\usepackage{natbib}

\shorttitle{Semi-implicit subgrid turbulence model}
\shortauthors{Semenov}
\graphicspath{{./}{figures/}}

\input{citation_fix}
\input{defs.tex}

\newcommand{\newtext}{}
\newcommand{\newtextb}{}

\begin{document}

\title{Capturing Turbulence with Numerical Dissipation:\\a Simple Dynamical Model for Unresolved Turbulence in Hydrodynamic Simulations}

\author[0000-0002-6648-7136]{Vadim A. Semenov}
\altaffiliation{\href{mailto:vadim.semenov@cfa.harvard.edu}{vadim.semenov@cfa.harvard.edu}}
\affiliation{Center for Astrophysics $|$ Harvard \& Smithsonian, 60 Garden St., Cambridge, MA 02138, USA}

\begin{abstract}
Modeling unresolved turbulence in astrophysical gasdynamic simulations can improve the modeling of other subgrid processes dependent on the turbulent structure of gas: from flame propagation in the interiors of combusting white dwarfs to star formation and chemical reaction rates in the interstellar medium, and nonthermal pressure support of circum- and intergalactic gas. 
We present a simple method for modeling unresolved turbulence in hydrodynamic simulations via tracking its sourcing by local numerical dissipation and modeling its decay into heat. This method is physically justified by the generic property of turbulent flows that they dissipate kinetic energy at a rate set by the energy cascade rate from large scales, which is independent of fluid viscosity, regardless of its nature, be it physical or numerical. We calibrate and test our model against decaying supersonic turbulence simulations. Despite its simplicity, the model quantitatively reproduces multiple nontrivial features of the high-resolution turbulence run: the temporal evolution of the average small-scale turbulence, its dependence on spatial scale, and the slope and scatter of the local correlation between subgrid turbulent velocities, gas densities, and local compression rates. As an example of practical applications, we use our model in isolated galactic disk simulations to model locally variable star formation efficiency at the subresolution scale. In the supersonic, star-forming gas, the new model performs comparably to a more sophisticated model where the turbulent cascade is described by explicit subgrid terms. Our new model is straightforward to implement in many hydrodynamic codes used in galaxy simulations, as it utilizes already existing infrastructure to implicitly track the numerical dissipation in such codes.
\end{abstract}


\section{Introduction}

Turbulent flows are ubiquitous in astrophysics on a wide range of scales, from planet atmospheres and the interiors of stars to the gaseous halos of galaxy clusters and the intergalactic medium. In hydrodynamic simulations, it is often infeasible to resolve such turbulent flow down to the physical dissipation scale. As an example, the resolution of state-of-the-art simulations of galaxy formation and the interstellar medium (ISM) can reach $\lesssim$parsec scales, while the dissipation occurs on $\sim$AU scales, i.e., six orders of magnitude below the resolution scale. In such cases, the kinetic energy is dissipated on the resolution scale, while the information about the turbulent cascade on unresolved scales is lost.

At the same time, many of the relevant small-scale processes affecting global behavior strongly depend on the turbulent structure of gas on unresolved scales. Explicit modeling of unresolved turbulence can substantially improve the modeling of such subgrid processes in a wide range of astrophysical applications and phenomena \citep[see, e.g.,][for reviews]{miesch15,schmidt-review}: turbulent flame propagation through the white dwarf interior during supernova (SN) Type Ia combustion \citep[e.g.,][]{niemeyer95,reinecke02}, unresolved star formation and feedback in galaxy-scale simulations \citep{braun15,semenov16,kretschmer20}, turbulent amplification of magnetic fields in the ISM \citep[][]{liu22}, turbulent support of gas during the formation of first black holes \citep[][]{latif13}, and nonthermal pressure support of galaxy clusters \citep[][]{scannapieco08-turb,maier09}, circumgalactic medium \citep[][]{schmidt21}, and intergalactic gas \citep[][]{iapichino11}, among others.
In this paper, we will use galaxy formation simulations as a practical example. In addition to the above applications, in galaxy simulations, subgrid turbulence models can also be used to account for unresolved gas clumping in cooling, heating, and chemical reaction processes, as well as acceleration and transport of cosmic rays \citep[see, e.g., ][for recent reviews]{zweibel17,ruszkowski23}. 

The subgrid modeling of turbulence in the above applications is often based on the so-called ``Large-Eddy Simulation'' (LES) methodology \citep[see][for extensive reviews]{sagaut,garnier}. The general idea of such models relies on the concept of scale separation, where the turbulent flow is split into the large-scale, energy-containing part (large eddies), which is modeled by solving modified hydrodynamic equations, and the small-scale part, which is followed by a subgrid model that describes the interactions with the resolved flow. Mathematically, the equations of LES can be obtained by applying a spatial filter to hydrodynamic equations and assuming closure relations for the subgrid terms. Such an approach is widely used in simulations of terrestrial turbulent flows with various applications, from industry and aerospace engineering to urban planning and geophysics simulations. 

An alternative approach to such hydrodynamic LES is not to include any explicit subgrid turbulence models at all. Although at first glance, the absence of such subgrid models might appear as a severe limitation of such simulations, the past few decades of practical applications have shown that this is not the case: simulations of turbulent flows without explicit models can still produce realistic results on scales sufficiently larger than the resolution. This phenomenon has been explained via the concept of Implicit Large-Eddy Simulations (ILES), whereby the numerical diffusivity of the chosen hydrodynamic method is argued to implicitly approximate the integral effect of all unresolved scales \citep[e.g.,][see \citealt{grinstein} for a review]{boris92,sytine00,margolin02,grete23,malvadi23}.

The ILES hypothesis is based on the generic property of turbulent flows that the rate of dissipation of kinetic energy is independent of the viscosity of the fluid and instead is set by the rate of energy cascade from large scales. The viscosity controls the spatial scale at which dissipation occurs: at a lower viscosity, the turbulent cascade extends down to smaller scales where viscous terms become important \citep[e.g.,][]{kolmogorov41,pope}. Thus, as long as the numerical method satisfies generic properties of the underlying hydrodynamic equations---such as conservation, causality, and locality---it will ensure the correct rate of energy transfer from large to small scales, while the flow structure and numerical errors of the scheme will conspire on close-to-resolution scales, ensuring dissipation of all incoming kinetic energy at the correct rate (\citealt{boris92,grinstein}; see also Section~\ref{sec:discussion} for a more detailed discussion).
The kinetic energy in such simulations is dissipated at the correct rate but on the wrong scale, which is set by resolution and numerics instead of physical viscosity. 
Extensive numerical studies show that, in the hydrodynamic codes commonly used in astrophysical applications, numerical dissipation effects kick in on rather large scales, corresponding to $\sim 30$ cells, and become dominant on $\sim 5\text{--}10$ cell scales \citep[e.g.,][]{kitsionas09,federrath10,federrath11,kritsuk11,grete23,malvadi23}. 
The existing simulations carried out with such codes without explicit subgrid turbulence models can be interpreted as ILES.

In its ultimate form, the goal of an ideal subgrid turbulence model is to account for the effect of unresolved turbulent motions on the resolved flow and thereby ``fix'' the simulated flow, so that it approximates a simulation where such scales would be resolved \citep[e.g.,][]{sagaut,grinstein,garnier}. 
The goal of the model presented in this paper is less ambitious.
The astrophysics applications discussed above, in particular, improving the subgrid modeling of processes that depend on turbulent gas structure, only require that the subgrid turbulence model captures a realistic amount of small-scale turbulence energy and its local correlations with other local gas properties (e.g., density, velocity, temperature, etc.). This does not necessarily require ``fixing'' the flow at the resolution scale as an ideal subgrid model would do. For example, in galaxy formation simulations, achieving such an ideal model is virtually impractical due to the subgrid processes, such as star formation and feedback, that are tightly coupled with unresolved turbulence but strongly affect global galaxy evolution. 
However, the modeling of these processes can be significantly improved by accounting for the effects of unresolved turbulence.

To this aim, we propose a hybrid method that combines features of both explicit and implicit LES, which we dub semi-implicit LES (SLES). In this method, the small-scale turbulent energy is followed explicitly, as is the case with standard LES. However, the turbulence source term is computed implicitly by capturing locally dissipated resolved kinetic energy and converting it into subgrid turbulence. This is motivated by the above feature of turbulent flows that the local dissipation rate is set by the energy cascade rate, which constitutes the basis of ILES.
The model is straightforward to implement as it relies on the existing infrastructure of ILES codes, while also making the local estimate of unresolved turbulence readily available, as in explicit LES.

The paper is structured as follows. In Section~\ref{sec:model}, we outline further details of the subgrid turbulence models and describe our implementation. Section~\ref{sec:turb} presents the calibration and tests of the model in a decaying supersonic turbulence setup. Section~\ref{sec:galaxy} presents the results of the model applied in a galaxy formation simulation. Section~\ref{sec:discussion} discusses our results, and Section~\ref{sec:summary} summarizes our conclusions.
In addition, Appendix~\ref{app:dns-convergence} investigates the convergence of our direct high-resolution simulation of decaying turbulence that we used to calibrate and test our model.

\section{Modeling unresolved turbulence}
\label{sec:model}

\begin{figure*}
\centering
\includegraphics[width=\textwidth]{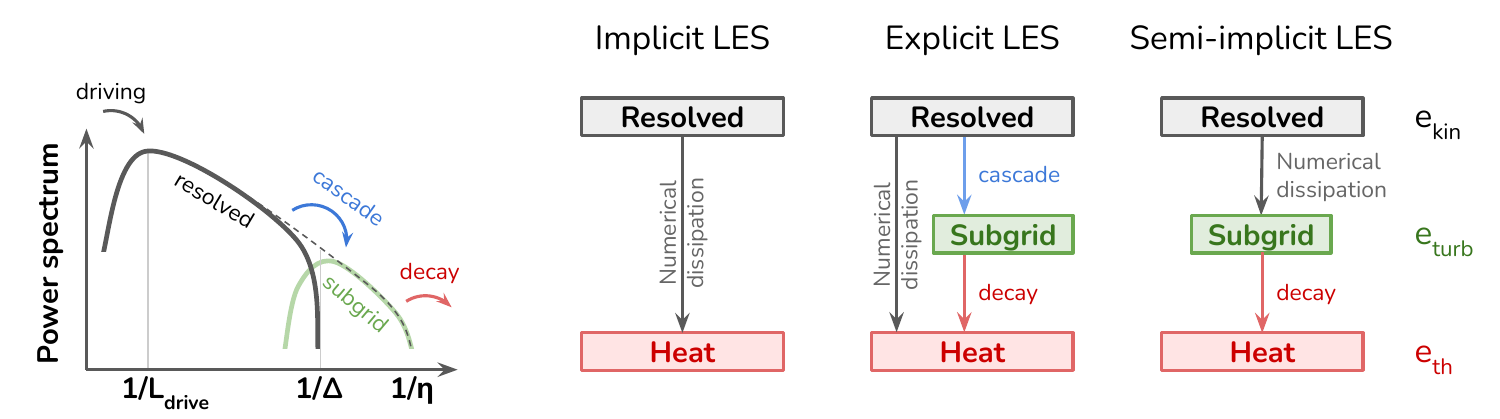}
\caption{\label{fig:schematics} Schematic overview of different types of subgrid turbulence models: implicit (ILES), explicit (LES), and proposed in this work, semi-implicit Large-Eddy Simulations (SLES). The SLES model combines the explicit modeling of the subgrid turbulent energy and its decay, similar to explicit LES, with the implicit source term provided by numerical dissipation, akin to ILES. The labels on the $x$-axis of the power spectrum cartoon represent the driving scale ($L_{\rm drive}$), resolution scale ($\Delta$), and physical dissipation scale ($\eta$). Green colors illustrate the explicit subgrid turbulence energy, $\eturb$, while ``cascade'' and ``decay'' represent the source and sink terms $\Sigma$ and $\epsilon$ in Equation~(\ref{eq:eturb}), respectively.}
\end{figure*}

In simulations of turbulent astrophysical flows, it is often infeasible to resolve the physical dissipation scale. Instead, the kinetic energy is dissipated on the scales set by the resolution, typically corresponding to $\sim$5--10 cells depending on the accuracy of the solver, with numerical effects kicking in on even larger scales, $\sim$30 cells \citep[e.g.,][]{kitsionas09,kritsuk11}. The information about the turbulent cascade on smaller scales is lost, as illustrated in the left panel of Figure~\ref{fig:schematics}. The goal of a subgrid turbulence model is, generally, to predict the properties of the missing small-scale turbulence, indicated in the cartoon with the green color. 

Below, we briefly review two existing broad types of methods used to model subgrid turbulence, implicit and explicit LES, and introduce our new hybrid method that we dub ``semi-implicit LES'' (SLES), as it combines the elements from both.
Note that all these models rely on the idea of a local kinetic energy cascade continuous in spatial scales. In general, the notion of cascade becomes nuanced for supersonic turbulence due to compressibility effects, but it is still applicable in simulations of isothermal supersonic turbulence, particularly in the context of galaxy formation and ISM simulations (see Section~\ref{sec:discussion}).

As a proof of concept for our new model, we have implemented it in the adaptive mesh refinement hydrodynamics and $N$-body code Adaptive Refinement Tree \citep[ART;][]{kravtsov99,kravtsov02,rudd08,gnedin11}, alongside the previously existing explicit LES model \citep{semenov16}. 
The hydrodynamic fluxes in the ART code are handled by a second-order Godunov-type method \citep{colella-glaz} with a piecewise linear reconstruction of states at the cell interfaces \citep{vanleer79} and a monotonized central slope limiter based on \citet{colella85}, with a \citet{lapidus67} artificial viscosity added in regions with strong density gradients. Adiabatic evolution of subgrid turbulent and thermal energies is followed using the entropy-conserving method described in \citet{semenov22}, while all extra source and sink terms are added in an operator-split manner. 
The following subsections provide additional details of our implementation of different subgrid turbulence models.

\subsection{Implicit Large-Eddy Simulation (ILES)}
\label{sec:model:iles}

The simplest case is the lack of any explicit model for unresolved scales at all. Only the resolved motions are modeled by solving equations of formally inviscid hydrodynamics, and these resolved motions are dissipated into heat at the resolution scale by numerical viscosity. This case is illustrated in the first panel on the right side of Figure~\ref{fig:schematics}.

Even though the viscosity in such simulations is purely numerical in nature, it ensures the physically correct dissipation rate, equal to the energy cascade rate through the resolved scales. 
This follows from the generic property of turbulent flows that the dissipation rate does not depend on the magnitude or even the nature of viscosity: the kinetic energy is dissipated by cascading down from large to small scales, while viscosity only sets the smallest scale on which these motions are dissipated locally. Thus, in these simulations, the dissipation occurs on the wrong scale (set by the resolution and solver details) but at the correct physical rate at which that energy would be transferred to unresolved scales if such scales were present. In this sense, the effects of numerical viscosity on the resolved flow can be viewed as the integral effect of unresolved scales, implying that an effective subgrid-scale model is implicitly built into the hydrodynamic method. Such an approach to modeling unresolved turbulence is collectively termed Implicit Large-Eddy Simulations\footnote{This term is derived from the class of methods with explicit modeling of subgrid scales, ``Large-Eddy Simulations,'' which is explained in the next section.} \citep[ILES; see][for an extensive review]{grinstein}. 

For this concept to work, the hydrodynamic scheme must respect the basic physical properties of the underlying hydrodynamic equations being solved, such as conservation, causality, positivity of positive-definite quantities (e.g., mass and energy), and monotonicity (lack of unphysical oscillations). Conservation, in particular, ensures that all dissipated kinetic energy instantly appears as heat as it bypasses the cascade through all unresolved scales. Therefore, in principle, thermal energy in simulations without explicit subgrid turbulence and no radiative cooling can be interpreted as a combination of thermal and subgrid turbulent motions but practical applications do not make such a distinction. In particular, in simulations of galaxy formation and ISM, this energy is subject to radiative cooling and can be quickly radiated away at the incorrect thermal cooling rate.

Implementation-wise, a simulation without any explicit subgrid terms can be interpreted as ILES. For example, the direct high-resolution simulation of decaying turbulence that we use as a reference in Section~\ref{sec:turb} is an ILES.
However, this method cannot be used for the purposes outlined in the Introduction---improving the subgrid modeling of processes coupled with turbulence on unresolved scales---as it does not predict the behavior of turbulence on unresolved scales but rather encapsulates its effects on the resolved flow implicitly.

\subsection{Explicit Large-Eddy Simulation (LES)}
\label{sec:model:les}

Another approach to modeling unresolved turbulence is by explicitly including the terms describing the interactions between subgrid and resolved scales in the hydrodynamic equations. Such a class of methods is called Large-Eddy Simulations and it is particularly widespread in various applications of simulations of terrestrial turbulence, such as aerospace engineering, urban planning, and climate modeling, among others \citep[see][for reviews]{sagaut,garnier}. Such explicit models are also used in astrophysical applications (see the Introduction and \citealt{miesch15} and \citealt{schmidt-review} for reviews), although in, e.g., galaxy formation simulations, they are not as widespread \citep[see][as examples]{schmidt14,semenov16,kretschmer20}.

The mathematical framework of LES is based on applying a spatial filter representing the finite resolution of a simulation to the hydrodynamic equations, which leads to a new, modified set of equations describing the evolution of unresolved turbulence as well as the extra terms in the original equations that describe the interaction between the subgrid scales and the resolved flow. To make the modified system of equations solvable, such terms require closure relations that cannot be derived from the first principles and instead are selected empirically or based on physical considerations and often are calibrated against high-resolution direct simulations. One of the key assumptions in such models is that the coupling between subgrid and resolved scales is facilitated by the largest unresolved eddies (which is the origin of the term LES) and therefore the transport coefficients in the subgrid closures (e.g., viscosity and diffusion coefficients) are set by the characteristic velocity of such eddies, $\sim \sqrt{ 2\eturb / \rho}$.

This case is illustrated in the middle of Figure~\ref{fig:schematics}. One of the key terms that require a closure relation is the sourcing of subgrid turbulence via the energy cascade from the resolved scales (shown with the blue arrow in the sketch). Such a term acts akin to viscosity: it draws kinetic energy from the resolved motions and transfers it to subgrid turbulence. When this turbulent viscosity is added explicitly, it acts on the flow in parallel with the numerical viscosity, and in principle one needs to make sure that the latter is subdominant, e.g., by selecting a high-order method. In practice, however, if the main goal is to predict the local properties of subgrid turbulence rather than accurately model its backreaction on the flow, such a model can still be used with more diffusive, second-order accurate solvers typical for galaxy formation simulations \citep[e.g.,][]{semenov16,kretschmer20}. In this case, the local flow quantities that enter closure relations are affected by the resolution; however, this approach is still physically motivated because such resolution effects can be viewed as an implicit model for unresolved scales (see Section~\ref{sec:model:iles} and \citealt{kretschmer20}).

In the ART code, we implement an explicit LES model following \citet{schmidt14} as described in \citet{semenov16}. To this end, we add explicit subgrid terms in the hydrodynamic equations and follow an additional energy variable that describes the total subgrid turbulent energy in each cell, $\eturb$:
\begin{equation}
\label{eq:eturb} 
\begin{split}
    \frac{\partial \eturb}{\partial t} &+ \divg (\eturb \vect{u}) = -\Pturb \divg \vect{u} + \\
    &+ \Sigma - \epsilon + \divg (\kappa \grad \eturb ) + S.
\end{split}
\end{equation}

The first line describes purely adiabatic behavior (advection on the left and $PdV$ work on the right), which is directly analogous to thermal energy. The turbulent pressure is related to $\eturb$ via an ideal gas-like equation of state, $\Pturb = 2/3\; \eturb$, which corresponds to the same adiabatic index as that of the thermal gas, $\gamma = 5/3$, because unresolved eddies behave analogously to thermal motions. This nonthermal pressure acts on the gas and enters equations for momentum and total energy together with the thermal pressure. We solve the adiabatic part by using an entropy-conservative scheme described in \citet{semenov22}.

The second line in the equation contains the additional terms that describe the evolution of subgrid turbulence: from left to right, sourcing by cascade from the resolved scales ($\Sigma$), decay into heat ($\epsilon$), turbulence diffusion ($\divg \kappa \grad \eturb $), and extra explicit sources such as direct injection by stellar feedback ($S$). In the rest of the section, we discuss the source term $\Sigma$, while the other terms are described in Section~\ref{sec:model:decay}.

The source term describing the energy cascade, $\Sigma$, is the key element of the model as it characterizes the interaction between subgrid and resolved scales. Corresponding terms enter the equations for momentum and total energy as viscosity, thereby maintaining the conservative form. To preserve conservation, we apply these terms by modifying momentum and energy fluxes and compute the amount of energy that needs to be added to $\eturb$ (i.e., the $\Sigma$ term itself) from the difference between total and kinetic energy arising from these extra fluxes. Note that the latter difference can be reliably computed only when $\eturb$ is not too small compared to $\etot$ and therefore, when $\eturb < 10^{-3}\; \etot$, we instead compute $\Sigma$ explicitly and add it to $\eturb$ in an operator-split manner.

In the tests presented below, we use two different closure relations for $\Sigma$. In the supersonic turbulence test (Section~\ref{sec:turb}), we use the nonlinear closure proposed and calibrated in the supersonic regime by \citet[][]{schmidt11}; specifically, we use their Equation~(8) with $C_1 = 0.02$ and $C_2 = 0.7$. For the galaxy formation tests (Section~\ref{sec:galaxy}), we use the standard eddy-viscosity closure (specifically, the same equation but with $C_1 = 0.095$ and $C_2 = 0$) because we find that this viscous source term is important in the warm and diffuse transonic ISM, while, in the cold and dense supersonic ISM, subgrid turbulence is mainly controlled by the interplay between adiabatic compression and decay making it only weakly sensitive to the specific choice of the closure \citep[][see also Section~\ref{sec:discussion} below]{semenov16}. 

Finally, it is worthwhile to point out the connection with instantaneous models that estimate subgrid velocity dispersion from the gradients of the resolved velocity field and have become popular in recent galaxy formation simulations \citep[e.g.,][]{hopkins13-sfrecipe,trebitsch17,marinacci19,vintergatan1}. Such a model can be viewed as a stationary limit of LES where time-dependent terms are set to zero (i.e., the left-hand side and diffusion term in Equation~(\ref{eq:eturb})) and the level of subgrid turbulence is set by the balance between production and dissipation: $-\Pturb \divg \vect{u} + \Sigma - \epsilon + S = 0$ \citep[see also][]{kretschmer20}. We leave a detailed comparison with such models for future work.

\subsection{Semi-implicit Large-Eddy Simulation (SLES)}
\label{sec:model:sles}

In this paper, we propose a model that combines features of both implicit and explicit LES described above. Specifically, we propose using the local numerical dissipation rate---which provides a reasonable approximation for the turbulent cascade rate (see Section~\ref{sec:model:iles})---as a source term for the explicitly followed subgrid turbulent energy. Such a model can be thought of as an explicit LES model where the source term $\Sigma$ in Equation~(\ref{eq:eturb}) is computed implicitly, analogously to ILES. Conversely, this model can also be thought of as an ILES, where subgrid turbulence and thermal energies are followed separately and are subject to their own explicit sink and source terms. For these reasons, we dub this model Semi-implicit LES (SLES) to highlight the analogy with both explicit and implicit LES.

The advantage of such a model over an explicit LES is that it does not require assuming and implementing closure relations for viscous subgrid terms. It only requires computing the local dissipation rate, which can be done straightforwardly as described below. At the same time, unlike ILES, the model explicitly tracks the evolution of subgrid turbulent energy that can be used in the subgrid models for other processes that are sensitive to unresolved turbulence.

Our implementation of SLES in the ART code is closely related to the dual-energy formalism that was introduced in the early galaxy formation simulations to enable the estimation of gas temperature and pressure in highly supersonic regions \citep{ryu93,bryan95}. In such regions, gas thermal energy (which is required to compute $T$ and $P$) cannot be computed as the difference between the total and kinetic energy because $\etot \approx \ekin$, and therefore their difference is contaminated by numerical errors. The proposed solution was to track the thermal energy or entropy using a separate variable and switch between this variable and $\eth \equiv \etot - \ekin$ when appropriate. 

When additional nonthermal energy like subgrid $\eturb$ is added to the system, one needs to choose how to partition the difference $\etot - \ekin$ between $\eturb$ and $\eth$ \citep[][see also \citealt{gupta21} for this problem in the context of cosmic-ray hydrodynamics]{semenov22}. In particular, the difference $\etot - \ekin$ implicitly contains the adiabatic change of $\eturb$ and $\eth$ as well as the kinetic energy dissipated locally during the hydrodynamic step, and therefore, this choice will dictate how this dissipated energy is split between $\eturb$ and $\eth$. To put all dissipated energy into $\eturb$ as envisioned in our simple SLES model, one simply needs to update $\eturb$ after the hydrodynamic step as
\begin{equation}
\label{eq:eturb-implicit-source} 
\eturb = \etot - \frac{(\rho \vect{u})^2}{2 \rho} - e_{\rm th,adi},
\end{equation}
where $e_{\rm th,adi}$ is the solution for the adiabatic part of the equation for $\eth$ solved akin to the dual-energy formalism. This step will produce an implicit solution for the adiabatic evolution of $\eturb$ and the source term due to the cascade of the dissipated resolved kinetic energy (i.e., the first line and $\Sigma$ in Equation~\ref{eq:eturb}). After that, all other explicit terms can be added as in LES.
It is worth noting, however, that kinetic energy can be dissipated into heat directly on shocks, which could be accounted for in the model by modifying this synchronization step (see Section~\ref{sec:discussion} for further discussion). 

In principle, one could not follow $\eturb$ explicitly but still apply all relevant subgrid terms (except for turbulent viscosity) to other quantities, computing $\eturb$ from Equation~(\ref{eq:eturb-implicit-source}) when needed. However, it is advantageous to follow $\eturb$ explicitly so that it can be followed even in regimes where the right-hand side of Equation~(\ref{eq:eturb-implicit-source}) is negative. Avoiding such an unphysical behavior requires enforcing local entropy stability of the numerical scheme (i.e., entropy must only increase), which is generally not guaranteed (see also Section~\ref{sec:discussion}). Instead, in our proof-of-concept implementation, we simply turn off this implicit source term in regions with negative dissipation. We find that only a small fraction of cells is affected in our turbulence tests below, and the overall results remain the same.

Finally, it is worth noting that the above model does not affect the \emph{adiabatic} evolution of gas. Indeed, only the fact that the residual energy ($\etot - \ekin$) is partitioned between $\eth$ and $\eturb$ distinguishes it from ILES; however, given that both energies obey the same $\gamma = 5/3$ equation of state, the pressure exerted on the gas remains exactly the same. This is also true when the dissipation term ($\epsilon$ in Equation~(\ref{eq:eturb})) is modeled explicitly, as it only changes the partitioning between $\eth$ and $\eturb$ but not their spatial distribution. This implies that the performance of the code in all idealized adiabatic tests without radiative cooling, like a shock tube or a point-source explosion, remains exactly the same as without this model.
In the presence of cooling, on the other hand, the solution becomes different, as the model helps to retain some of the dissipated kinetic energy in the form of $\eturb$, which decays on a timescale different from the thermal cooling timescale.

\subsection{Turbulence Decay, Diffusion, and Extra Sources}
\label{sec:model:decay}

Other than $\Sigma$, the rest of the explicit source terms in the second line of Equation~(\ref{eq:eturb}) are the same in LES and SLES. Here we describe the adopted closures for these terms.

Subgrid turbulence dissipation into heat is parameterized via the decay time that is assumed to be close to the local cell-crossing time of the largest unresolved eddies, $t_{\rm cross} \equiv \Delta/(2\sigma) = \sqrt{\rho} \Delta /(2\sqrt{2 \eturb})$:
\begin{equation}
\label{eq:ce}
    \epsilon \equiv \ceprime \frac{\eturb}{t_{\rm cross}} = \ce \frac{(\eturb)^{3/2}}{\sqrt{\rho} \Delta},
\end{equation}
where $\ce \equiv 2\sqrt{2} \,\ceprime$ is an order-of-unity constant that needs to be calibrated.\footnote{This notation follows our implementation in the ART code, where the numerical factor $2\sqrt{2}$ is absorbed into the parameter $\ce$.} In our simple SLES model, this is the only tunable parameter that we calibrate against a high-resolution turbulent box simulation in Section~\ref{sec:turb:calib}.

The same term enters the equation for thermal energy (which is analogous to Equation~(\ref{eq:eturb})) but with a positive sign, implying that the dissipated subgrid turbulence appears as heat, which is then subject to radiative losses. 
The amount of dissipated subgrid turbulence, $\Delta \eturb$, during a time step, $\tau$, is calculated from the analytic solution of $d \eturb / d t = - \epsilon$:
\begin{align}
    \Delta \eturb &= \eturb \, \frac{(f+4)\,f}{(f+2)^2},\\
    f &\equiv \ceprime \, \frac{\tau}{t_{\rm cross}} = \ce \frac{\tau}{\Delta} \sqrt{ \frac{\eturb}{\rho} },
\end{align}
with $\Delta \eturb$ subtracted from $\eturb$ and added to $\eth$ after each step.

Next, for the turbulent diffusion term ($\divg \kappa \grad \eturb$), we adopt turbulent diffusivity of $\kappa \equiv C^\prime_\kappa \sigma \, \Delta= C_\kappa \sqrt{\eturb/\rho} \Delta$ with $C_\kappa = \sqrt{2} \, C^\prime_\kappa = 0.4$ \citep{schmidt14,schmidt}. A similar term is applied to all advected scalars, including thermal energy and chemical abundances, when the latter are followed explicitly. In our tests, the effect of explicit subgrid turbulent diffusion is generally smaller than numerical diffusivity except for the regions with very high $\sigma$ such as the interiors of SN bubbles in galaxy simulations (Section~\ref{sec:galaxy}). The effect of turbulent diffusion is significantly more important in Lagrangian and quasi-Lagrangian (moving-mesh) codes owing to their reduced advection errors \citep[e.g.,][]{rennehan21}. 
As the choice of $C_\kappa$ does not affect our results, we do not consider it as a tunable parameter of the model, even though we keep the diffusion term on in the tests presented below. Turning off this term completely affects our results only weakly.

Finally, the explicit source term $S$ can be used to source subgrid turbulence via various unresolved processes, including stellar feedback. For example, a fraction of SN energy injected during a time step can be added directly to $\eturb$ via this term. In our galaxy simulations presented in Section~\ref{sec:galaxy}, we do not perform such an explicit injection (i.e., $S \equiv 0$), but we keep this term in Equation~(\ref{eq:eturb}) for completeness.

\section{Decaying supersonic turbulence}
\label{sec:turb}

In this section, we use a decaying supersonic isothermal turbulence test to calibrate and investigate the performance of our new SLES model. To this end, we use a high-resolution simulation to predict the distribution of turbulent energy on small scales (Figure~\ref{fig:turbbox-maps}, top row). This distribution is then compared with low-resolution resimulations of the same initial conditions where the small-scale turbulence is predicted using the SLES model (Figure~\ref{fig:turbbox-maps}, bottom row). As we will show, despite its simplicity, the SLES model is able to capture multiple nontrivial features of the direct numerical simulation (DNS), such as the time evolution of the mean turbulent energy, its dependence on scale, and the correlation between gas density and turbulent velocity.

\subsection{Numerical Setup and Analysis}
\label{sec:turb:setup}

\begin{figure*}
\centering
\includegraphics[width=\textwidth]{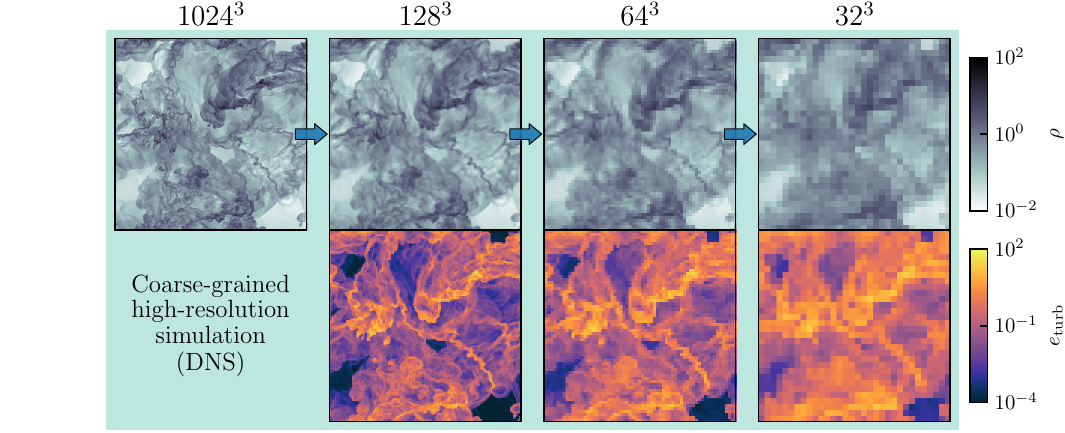}\\
\includegraphics[width=\textwidth]{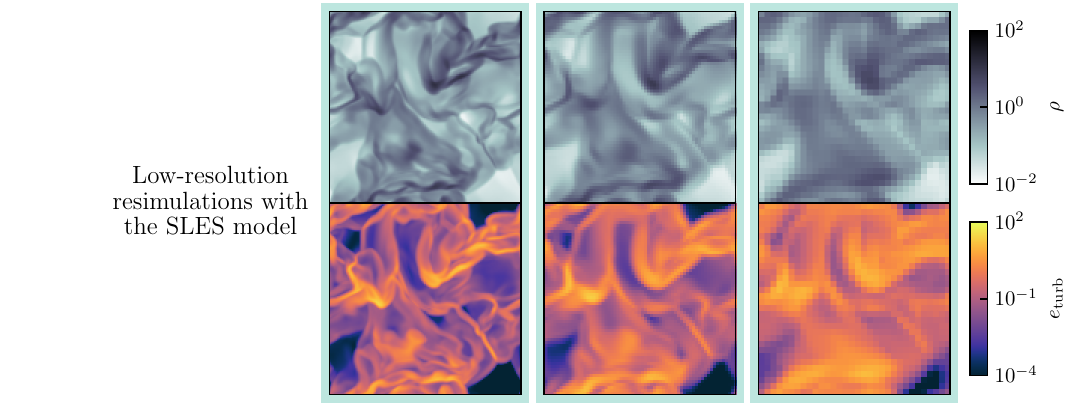}
\caption{\label{fig:turbbox-maps} Gas density (gray) and small-scale turbulent energy slices (orange) in the coarse-grained high-resolution simulation (DNS; the top set of panels) and low-resolution resimulations of the same ICs with the SLES model (the bottom set of panels). The maps are shown at $t = 2/3\;t_{\rm cross,0}$. The leftmost panel in the top row shows the density slice at the native resolution of the DNS, while the other columns show coarse-grained values, with density being volume-weighted and $\eturb$ computed from Equation~(\ref{eq:eturb-coarse}); the corresponding grid size is indicated at the top. The bottom panels show slices at the same time step from low-resolution resimulations using the corresponding grid sizes, where $\eturb$ is predicted using the SLES model. Thus, in total, four simulations are shown in the plot as indicated by the green shaded background: one DNS and three SLES resimulations. Low-resolution SLES resimulations produce smoother solutions than the DNS, but they capture the main global features of the small-scale turbulence, in particular, the correlation with gas density.}
\end{figure*}

To initialize the decaying isothermal turbulence test, we use the initial conditions from the code comparison project of \citet{kritsuk11}. The initial sonic Mach number in this setup is $\Mach \approx 9$, and it decreases with time as turbulence is allowed to decay. The results below assume units in which the box size is $\Lbox = 1$, the average density is $\langle \rho \rangle = 1$, and the constant sound speed is $\cs = 1$.

After the start, the simulation undergoes a brief initial transient stage that lasts approximately one turbulent box-crossing time (see the slight bump in the dashed lines at $t < t_{\rm cross,0}$ in Figures~\ref{fig:ce-calib} and \ref{fig:var-scale} below). This transient is mainly caused by two factors. First, the initial conditions were created using a higher-order method (piecewise parabolic method) than we employ in our run. Second, the original simulation included magnetic pressure, which we removed in our runs. This pressure is, however, dynamically subdominant (corresponding to the Alfv\'{e}n Mach number of $\approx 4.4$). Overall, the strength of the transient is rather small, and we present our analyses at sufficiently late times, $> t_{\rm cross,0}$, after it is dissipated.

As a reference for calibration and testing of the model, we run a high-resolution version of this setup with a $1024^3$ uniform grid. We do not model physical viscosity; instead, kinetic energy is dissipated numerically on scales close to the resolution. However, for comparisons with the SLES model, we compute the turbulent energy on scales larger than these numerical dissipation scales, on which the mean turbulent energy converges ($L \geq \Lbox/64$; see Appendix~\ref{app:dns-convergence}. For this reason, we henceforth refer to this reference simulation as a ``direct numerical simulation'' (DNS) even though, formally, the physical dissipation is absent in this simulation (the so-called ``Coarse DNS'' approach).

The distribution of small-scale turbulence is derived by coarse-graining the direct simulation and averaging the properties of the flow on the scales below the coarse-graining scale. To this end, we use uniform $N^3$ grids, where $N < 1024$ and is a power of 2 such that each coarse cell is a cubic volume with $1024/N$ cells on a side. To respect conservation laws, the averages of all conservative quantities are volume-weighted. For example, the coarse-grained density is computed as

\begin{equation}
\label{eq:rho}
    \langle \rho \rangle_{N} \equiv \frac{1}{dV_{N}} \sum_i \rho_i dV = \left( \frac{N}{1024} \right)^3 \sum_i \rho_i,
\end{equation}
where $dV \equiv (\Lbox/1024)^3$ and $dV_N \equiv (\Lbox/N)^3$ are the cell volumes of the original $1024^3$ and coarse-grained $N^3$ grids.

The coarse-grained velocity is not a conservative quantity, and for consistency with momentum conservation, it must be a mass-weighted average:\footnote{This procedure also corresponds to the so-called Favre averaging that is assumed in the mathematical formulation of the LES model used in our comparisons (see Section~\ref{sec:model:les}). Using this averaging helps to preserve the form of the mass continuity equation and avoid extra terms in the equations for energy and momentum \citep[e.g.,][]{garnier}.} 

\begin{equation}
    \vect{u}_{N} \equiv \frac{\langle \rho \vect{u} \rangle_{N}}{\langle \rho \rangle_{N}}.
\end{equation}
Correspondingly, the small-scale turbulent energy below the coarse-graining scale is computed as the difference between the total kinetic energy and the bulk coarse-grained kinetic energy:

\begin{equation}
\label{eq:eturb-coarse}
    e_{ {\rm turb,}N} \equiv \left\langle \frac{\rho |\vect{u}|^2}{2} \right\rangle_N - \frac{\langle \rho \rangle_{N} \, |\vect{u}_{N}|^2}{2}.
\end{equation}
To simplify the notation, we will omit the brackets and the index showing the filtering grid size $N$; thus, in the context of coarse-grained DNS, $\rho$, $\vect{u}$, and $\eturb$ will refer to the averaged quantities defined in Equations~(\ref{eq:rho})--(\ref{eq:eturb-coarse}).

Implementation-wise, we perform the above averaging by exploiting the nested oct structure of the computational mesh in the ART code. Specifically, we use a static $16^3$ root grid where each cell is split into octs until the target uniform resolution of $1024^3$ is achieved, which corresponds to 7 refinement levels. Then, starting from the highest refinement level, hydrodynamic quantities are averaged between the cells in a given oct and saved level by level, providing a hierarchy of coarse-grained versions of the direct simulation. 

The top row of Figure~\ref{fig:turbbox-maps} shows the original high-resolution density map (the leftmost panel) and the series of coarse-grained versions using uniform grids with $N=128$, 64, and 32 cells on a side at $t = 2/3\;t_{\rm cross,0}$, where $t_{\rm cross,0}$ is the initial turbulent box-crossing time.\footnote{The specific time at which the maps are shown is picked such that the small-scale distribution of $\eturb$ has time to develop, but the initial large-scale turbulent energy has not dissipated away.} The top (gray) and bottom (orange) subsets of maps show, respectively, the coarse-grained density (Equation~(\ref{eq:rho})) and the turbulent energy on the scales below the coarse-graining scale ($\eturb$; Equation~(\ref{eq:eturb-coarse})). The distribution of $\eturb$ shows a clear correlation with density: the higher turbulent energy corresponds to denser regions. In addition, the global mean value of $\eturb$ increases with decreasing resolution of the coarse-graining grid, i.e., the map becomes more orange from left to right. This is because larger scales contain more turbulent energy. 

The aim of a subgrid turbulence model is to predict the distribution of $\eturb$ in a low-resolution run. 
To this end, we rerun our simulation at low resolutions starting from the coarse-grained version of the same initial conditions as in the DNS and turning on the SLES model to track $\eturb$ on the fly. 

The results of such low-resolution resimulations with the SLES model are shown in the bottom set of panels. 
These low-resolution resimulations reproduce the overall features of the direct simulation, such as the locations of the high-density structures and the correspondence between high-density and high-$\eturb$ regions. The solution is significantly smoother than in the DNS owing to higher numerical viscosity, i.e., larger numerical dissipation scales. The SLES model also captures the increase of the mean $\eturb$ at a lower resolution.

Having such a set of simulations, the SLES model can be calibrated and tested by comparing its predictions with the coarse-grained DNS. 
The cell-by-cell comparison of $\eturb$ is complicated due to the strong effect of resolution on the flow structure. Indeed, even minute changes due to, e.g., variations in the time-stepping scheme can cause slight offsets of the large-scale features like the locations of the shock fronts, leading to large cell-to-cell differences between the two runs (see, e.g., Figure~12 in \citealt{gnedin18}). For that reason, in the subsequent sections, we compare global statistics: the mean $\eturb$ averaged over the entire box (which we also use to calibrate the model in Section~\ref{sec:turb:calib}); the normalization, slope, and scatter of the relation between gas density and $\eturb$; and the power spectra of resolved kinetic energy, density, and $\eturb$.

\subsection{Model Calibration}
\label{sec:turb:calib}

\begin{figure}
\centering
\includegraphics[width=\columnwidth]{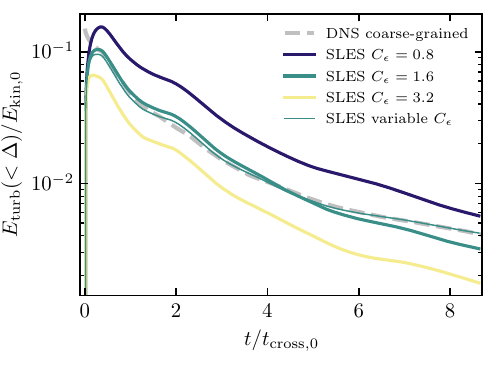}
\caption{\label{fig:ce-calib} Comparison of the global mean small-scale turbulent energy, $\Eturb \equiv \langle \eturb \rangle_{\rm box}$, in coarse-grained DNS (dashed line) and low-resolution SLES resimulations with different values of the dissipation parameter, $\ce$ (colored lines). For this comparison, we use the grid size of 32$^3$, i.e., the cell size $\Delta \equiv \Lbox/32$. In the DNS, $\eturb$ is computed following Equation~(\ref{eq:eturb-coarse}) while in the low-resolution resimulations, it is tracked by the SLES model. The mean $\Eturb$ is normalized to the total initial kinetic energy, i.e., $\Ekininit \equiv \Eturb(<\Lbox)$ at $t=0$. The time axis is normalized by the initial box-crossing time of turbulence, $\tcross \equiv 0.5\;\Lbox/\sigma_0$, or $\approx 0.058$ sound-crossing times for the initial Mach number of $\Mach \approx 8.6$. The value of $\ce = 1.6$ provides the best match to the global $\Eturb$ and therefore we will use it as our fiducial choice in the rest of the paper. The agreement with the DNS can be further improved by making $\ce$ dependent on the local subgrid Mach number, as shown with the thin line and described in the text.}
\end{figure}

The simplest version of the SLES model involves a single parameter that needs to be selected: $\ce$, describing the rate of subgrid turbulence dissipation in units of the local eddy cell-crossing time (Equation~(\ref{eq:ce})). We calibrate the value of $\ce$ by matching the mean subgrid turbulent energy, $\Eturb \equiv \langle \eturb \rangle_{\rm box}$.

Figure~\ref{fig:ce-calib} compares the evolution of $\Eturb$ normalized by the initial total kinetic energy in 32$^3$ resolution simulations with the SLES model assuming different values of $\ce$. $\Eturb$ quickly reaches a peak initially and then decreases with time as the turbulence in the box decays. The normalization of $\Eturb$ scales roughly linearly with $\ce$. The linear scaling indicates that, on average, subgrid turbulence is in a stationary regime, where the source term is locally in balance with the dissipation term that is proportional to $\ce$ (see Equation~\ref{eq:ce}).

For reference, the dashed line shows the small-scale turbulent energy in the direct high-resolution simulation (DNS) coarse-grained on the scale corresponding to the 32$^3$ grid. The SLES model with $\ce = 1.6$ provides a reasonably close match to the DNS results, and therefore we select this value as our fiducial choice. A close-to-unity value of $\ce$ is physically sensible as it implies that the local dissipation timescale of subgrid turbulence is close to the eddy cell-crossing time.

Note that as the model contains only a single parameter, only a single instantaneous value of $\Eturb$ from one of the time steps could be used for calibration. The fact that the SLES model also reasonably closely reproduces the temporal behavior of $\Eturb$ is nontrivial and can be viewed as a test of the SLES model rather than a part of calibration (see also Section~\ref{sec:turb:perf:Eturb}). Indeed, as the global Mach number of decaying turbulence is decreasing, the values of $\Eturb$ separated by more than the global turbulence crossing time, $\tcross$, can be interpreted as independent realizations of different Mach number regimes.

Finally, note that, although the simple SLES model captures the overall behavior of $\Eturb$ in the DNS, there are some differences in detail. In particular, around $t \sim 2\;\tcross$, the value of $\Eturb$ in the SLES model overshoots DNS by $\sim 20\%$ and then decays at a somewhat faster rate, resulting in $\sim 20\%$ lower $\Eturb$ than that of DNS at $t \gtrsim 8\;\tcross$. 

One of the reasons for such a difference can be the dependence of the subgrid turbulence dissipation rate on the local properties of the flow. Indeed, these results suggest that turbulence dissipation slows down with time as the turbulence decays and the crossing time increases. As an illustration, the thin line in the figure shows a modified version of the SLES model, where the local dissipation rate decreases in the subsonic regime by analogy with the global behavior of $\Eturb$. Specifically, we reduce the value of $\ce$ to 0.8 when the local Mach number of \emph{subgrid} turbulence becomes $\Mach_{\rm sgs} \equiv \sqrt{(2 \eturb / \rho)}/\cs \leq 0.5$, with a smooth transition to the fiducial $\ce=1.6$ at $\Mach_{\rm sgs} \geq 0.7$.

Such a modification helps to improve the fit. We stress, however, that although this modification is motivated by the qualitative behavior of $\Eturb$, the parameter choices are not based on any physical model; this model is shown here for illustration purposes only. In the rest of the paper, we will focus on the SLES model with a constant value of $\ce$. Given the simplicity of the model, the $\sim 20\%$ deviations of the mean $\eturb$ are acceptable, and we leave the investigation of more detailed models for $\ce$ to future work.

\subsection{Performance of the SLES Model}
\label{sec:turb:perf}

In this section, we focus on our fiducial SLES model with a constant dissipation rate parameter calibrated in the previous section, $\ce = 1.6$, and compare its predictions with the results of the coarse-grained high-resolution simulation (DNS).

\subsubsection{Mean Small-scale Turbulent Energy}
\label{sec:turb:perf:Eturb}

\begin{figure}
\centering
\includegraphics[width=\columnwidth]{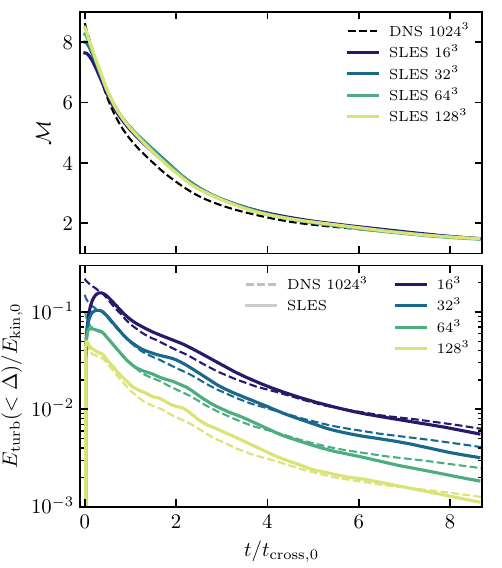}
\caption{\label{fig:var-scale} Comparison of coarse-grained DNS (dashed lines) with the SLES resimulations at corresponding resolutions (colored lines). The top panel shows the evolution of the global Mach number, $\Mach = \sqrt{ 2 E_{\rm kin} }$ in the adopted units ($\langle \rho \rangle \equiv 1$). The bottom panel shows the mean small-scale turbulent energy, $\Eturb \equiv \langle \eturb \rangle_{\rm box}$, normalized by the initial total kinetic energy in the box, $E_{\rm kin,0}$. The SLES model captures both the evolution of $\Eturb$ and its scale dependence. As shown in Appendix~\ref{app:dns-convergence}, $\Eturb$ is underestimated for 128$^3$ due to a lack of convergence on these small scales; however, we still show the results for this grid to demonstrate that the scaling of $\Eturb$ with the grid size remains the same in SLES.}
\end{figure}

As pointed out in the previous section, the fact that the SLES model reproduces the temporal evolution of the mean small-scale turbulent energy, $\Eturb$, in the DNS is a nontrivial test of the model, as the time steps spaced by more than a box-crossing time can be viewed as independent realizations. In decaying turbulence, the global Mach number of turbulence decreases with time, and therefore, different times also probe different Mach number regimes. As the top panel of Figure~\ref{fig:var-scale} shows, the Mach number decreases from $\sim 8.5$ to $\lesssim 2$ after 8 eddy crossing times of the simulation box, $\tcross$. The SLES model performs well in these different regimes.

Another test of the model is its performance on different spatial scales. In the previous section, the model was calibrated on a specific scale, $\Lbox/32$, corresponding to a $32^3$ grid. The bottom panel of Figure~\ref{fig:var-scale} shows the model's performance on different scales, corresponding to $16^3$, $32^3$, $64^3$, and $128^3$ grids. The dashed lines show the DNS coarse-grained on these scales, while the solid lines are low-resolution resimulations with the SLES model using corresponding grids.

The shape of $\Eturb$ evolution is qualitatively similar on different scales and is only offset to higher values for larger scales (i.e., smaller $N^3$). This is because larger scales contain more kinetic energy. The SLES model captures this dependence on scale. In this model, the information about scale is contained both in the implicit source term---as the numerical dissipation rate depends on the resolution---and in the explicit turbulence decay term, which is proportional to the local cell-crossing time of subgrid turbulence (Equation~(\ref{eq:ce})).

One noticeable deviation of the SLES resimulations from the DNS is the slightly higher global Mach number and mean $\Eturb$ at $t \sim 1\text{--}3 \times \tcross$. As was shown in the previous section, such a feature can be suppressed by modifying the model for the subgrid turbulence decay (see Figure~\ref{fig:ce-calib}). However, this feature might also be due to the initial transient associated with the smoothing of the initial sharp cell-to-cell density and velocity gradients resulting from the coarse-graining high-resolution ICs. We checked that this feature does not depend on how the subgrid $\eturb$ is initialized. In these tests we initialize as $\eturb = 0$ throughout the box; however, if we start from the $\eturb$ equal to the coarse-grained ICs values using Equation~(\ref{eq:eturb-coarse}), this feature still persists. Such a transient can be avoided by testing the model in a continuously driven turbulence box setup, which we leave to a separate study.

\subsubsection{Turbulent Velocity--Density Correlation}
\label{sec:turb:perf:n-sigma}

\begin{figure*}
\centering
\includegraphics[width=\textwidth]{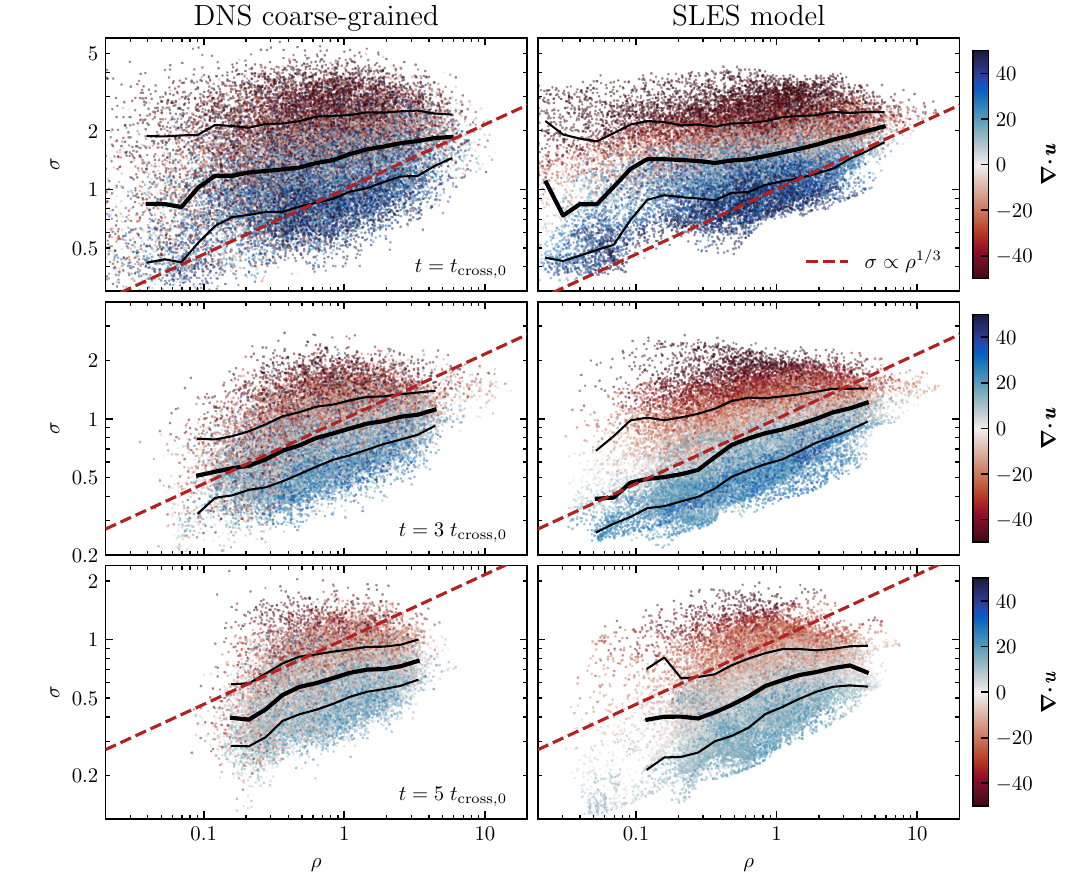}
\caption{\label{fig:n-sigma} Cell-by-cell correlation between gas density and small-scale velocity dispersion, $\sigma \equiv \sqrt{ 2 \eturb / \rho}$ on a 32$^3$ grid. The left column shows the DNS, i.e., each point shows coarse-grained values described in Section~\ref{sec:turb:setup}; the right column shows a 32$^3$ resimulation with the SLES model, where each point shows cell values at the native resolution, with $\eturb$ being predicted by the SLES model. The color shows the divergence of coarse-grain (for DNS) or resolved (for SLES) velocity. The rows correspond to different time steps: from top to bottom, 1, 3, and $5\;\tcross$. Solid black lines show the medians and 18$^\text{th}$ to 84$^\text{th}$ interpercentile ranges. The red dashed line shows the adiabatic scaling corresponding to $\eturb \propto \rho^{5/3}$, with the normalization being the same in all panels. The SLES model captures the relation between $\rho$ and $\sigma$ in terms of normalization, slope, the magnitude of scatter, and its scaling with the resolved velocity divergence. }
\end{figure*}

\begin{figure}
\centering
\includegraphics[width=\columnwidth]{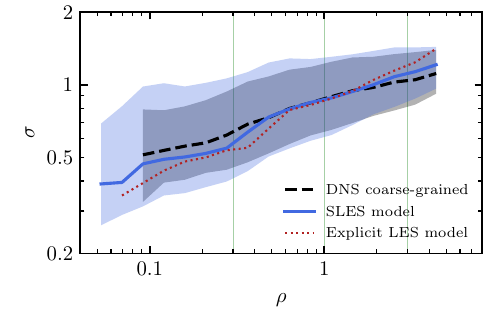}\\
\includegraphics[width=\columnwidth]{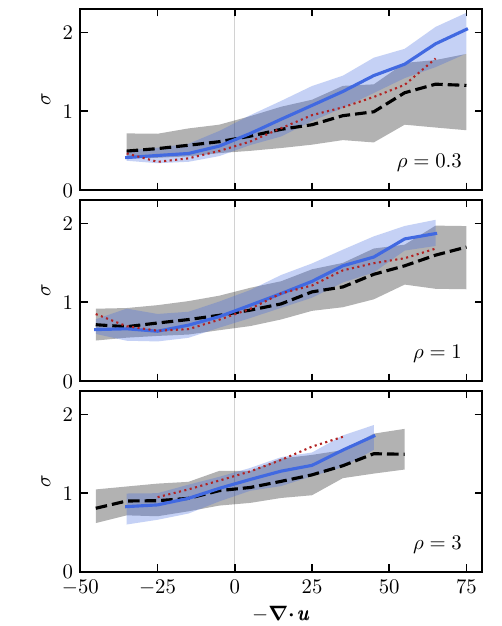}
\caption{\label{fig:nsigma-scatter} {\bf Top panel:} comparison of the $\rho$--$\sigma$ relation between the coarse-grained DNS (dashed black lines) and low-resolution SLES resimulation (solid blue lines). {\bf Bottom set of panels:} dependence of scatter in $\sigma$ on gas compression rate (negative divergence of the resolved velocity field, $-\divg\vect{u}$) at three fixed densities indicated with green lines in the top panel and labeled in the corner of each bottom panel. Each of the panels shows cells with densities bound by 1.5 times higher and 1.5 times lower than these values. For reference, the dotted red lines show the medians from the low-resolution resimulations with an explicit LES model proposed by \citet{schmidt11}. The SLES model qualitatively reproduces the trend of $\sigma$ with $\divg\vect{u}$ although it does show a steeper increase at high compression rates. This behavior, however, is analogous to that in a more complex explicit LES model.}
\end{figure}

As Figure~\ref{fig:turbbox-maps} shows, local turbulent energy on small scales is correlated with the gas density. Figure~\ref{fig:n-sigma} demonstrates this correlation quantitatively for both the DNS coarse-grained using a $32^3$ grid (left column) and an SLES resimulation at the corresponding resolution (right column). The rows correspond to different times: from top to bottom, 1, 3, and $5$ global turbulence crossing times of the simulation box, $\tcross$. The range of the $y$-axis changes from top to bottom to compensate for the decrease in the normalization of this relation due to turbulence decay; to help visualize this decreasing trend, the red line is kept the same in all panels (see below).

First, focusing on the DNS, the $\rho$--$\sigma$ relation exhibits several remarkable features both in slope and in scatter. 
The slope is shallower at earlier times and at higher $\sigma$, implying that turbulent energy changes isothermally with density. At later times and at lower $\sigma$, the slope steepens and becomes close to the adiabatic value, $\sigma \propto \rho^{1/3}$ (i.e., $\eturb = \rho \sigma^2 / 2 \propto \rho^{5/3}$), shown with the red thin line. This change is the most noticeable in the bottom panel, where the median $\st$ exhibits a change in slope at $\rho \sim 0.5$. Overall, the changes in the slope are subtle, indicating that, on average, small-scale turbulence remains in the intermediate regime. 

As the color shows, the scatter correlates with the divergence of resolved velocity, i.e., the gas compression rate. This reflects the adiabatic heating (cooling) of turbulence as contracting (expanding) gas exerts the $PdV$ work on subgrid turbulence \citep{robertson12}. The magnitude of the compression rate decreases with time as the turbulence decays, but the correlation of scatter with $\divg\vect{u}$ persists.   

A comparison of the right and left columns in the figure shows that the SLES model captures the slope, the magnitude of scatter, and its dependence on compression rate remarkably well. Such agreement is particularly remarkable given how simple the model is: it involves only one tunable parameter that is fixed effectively by matching the normalization of this relation in one of the snapshots (see Section~\ref{sec:turb:calib}). The slope, the scatter, and their evolution are the predictions of the model, and they match the DNS results.

To further quantify the agreement between the SLES model and DNS, the $\rho$--$\sigma$ relations from both simulations are overplotted in the top panel of Figure~\ref{fig:nsigma-scatter}. 
The bottom set of panels shows the dependence of $\sigma$ on the resolved compression rate at three fixed values of $\rho = 0.3$, 1, and 3. The SLES model captures the overall increase of $\sigma$ at higher compression rates, $-\divg\vect{u}$, although it produces a steeper relation at high values of $-\divg\vect{u}$. However, given the simplicity of the model, the agreement is still remarkable. For reference, the thin dotted line shows the results of a low-resolution simulation with an explicit LES model adopting the nonlinear \citet{schmidt11} closure for the supersonic turbulence source (see Section~\ref{sec:model:les}). The performance of the SLES model is comparable to that of the explicit LES model without requiring assumptions about the closure relation for the turbulence source term.

\subsubsection{Power Spectra}
\label{sec:turb:perf:Pk}

\begin{figure}
\centering
\includegraphics[width=\columnwidth]{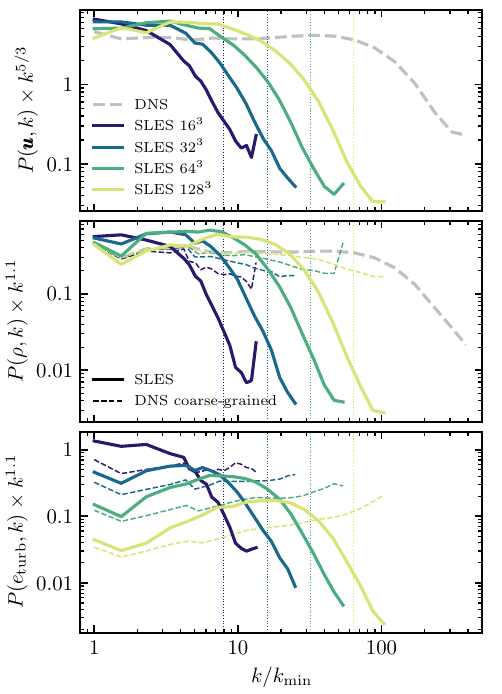}
\caption{\label{fig:power-spectra} Power spectra in DNS (dashed lines) and SLES resimulations (solid colored lines): velocity ($\vect{u}$; top), density ($\rho$; middle), and small-scale turbulent energy ($\eturb$; bottom). The velocity power spectrum is compensated by $k^{-5/3}$, while $\rho$ and $\eturb$ spectra are compensated by $k^{-1.1}$. Gray dashed lines show the results from DNS at native 1024$^3$ resolution, while thin colored dashed lines show coarse-grained statistics on the grids corresponding to the SLES resimulations. The wave number, $k$, is normalized by the minimal value corresponding to the box size, $k_{\rm min} = 2 \pi/\Lbox$. Vertical dotted lines show Nyquist frequencies for low-resolution grids, i.e., $k_{\rm Ny} = \pi \Ngrid/\Lbox$.}
\end{figure}

The $\rho$--$\sigma$ relation investigated in the previous section is a single-point statistic of the local properties of the flow. 
In this section, we investigate the power spectrum, which encodes the information about the scale dependence of the spatial distribution of gas properties. Figure~\ref{fig:power-spectra} shows the power spectra of resolved velocities ($\vect{u}$), gas density ($\rho$), and small-scale turbulent energy ($\eturb$).
The power spectra are calculated following Section~4.1 in \citet{kritsuk11}.

As the dashed line in the top panel shows, in the DNS, the velocity power spectrum is a power law with a constant slope of $\approx -5/3$ down to $k/k_{\rm min} \sim 100$ and falls off at higher $k$. Given the resolution of the DNS, $1024^3$, the falloff at these scales indicates that, in the ART code, numerical dissipation becomes important on the scales of $\sim 10$ grid cells. A similar numerical dissipation scale was found by \citet{kritsuk11} for analogous second-order hydrodynamic solvers \citep[e.g., the RAMSES code;][]{ramses}. It is worth noting that the power spectrum compensated by the $k^{-5/3}$ slope exhibits a local maximum (i.e., the bottleneck effect) at $k/k_{\rm min} \sim 40$ indicating that numerical dissipation effects kick in on $\sim1024/40\sim25$ cell scales \citep{kitsionas09,federrath10,federrath11}, which can also explain the excess of power at low $k$ in the SLES runs.

As the spatial resolution is degraded in SLES resimulations, the kinetic energy is dissipated on larger scales (see the colored lines in the top panel of Figure~\ref{fig:power-spectra}). For reference, the vertical dotted lines in the figure show the Nyquist wavenumbers corresponding to 2 grid cells at a given resolution: $k_{\rm Ny} = \pi \Ngrid/\Lbox$. 
The offset of numerical dissipation to larger and larger scales leads to the distribution of density becoming progressively smoother (see the bottom set of maps in Figure~\ref{fig:turbbox-maps}). 

As the middle panel of Figure~\ref{fig:power-spectra} shows, the corresponding falloff also appears in the density power spectrum. This falloff also shifts to larger scales (smaller $k$) with decreasing resolution. The density power spectrum falls off with $k$ roughly as $P(\rho,k) \propto k^{-1.1}$.
This slope is within the range reported by \citet{federrath13} for the nonmagnetized turbulence runs with Mach numbers $\lesssim 10$ before the gas self-gravity is turned on \citep[see also][]{federrath10}.

Finally, the bottom panel of Figure~\ref{fig:power-spectra} compares the power spectra of local small-scale turbulence, $\eturb$, compensated by $k^{-1.1}$ to simplify the comparison with the density power spectra. The slope of the $\eturb$ power spectrum is somewhat shallower than that of density, reflecting deviations from the isothermal (i.e., $\eturb \propto \rho$) behavior on small scales (see Section~\ref{sec:turb:perf:n-sigma}). The increase of the magnitude of the $\eturb$ power spectrum at lower resolution again reflects that larger scales contain more turbulent energy (compare with the bottom panel of Figure~\ref{fig:var-scale}). 

Apart from having qualitatively similar slopes, the drop-off of the $\rho$ and $\eturb$ power spectra at high $k$ exhibits similar dependences on the resolution. This is another manifestation of the $\rho$--$\sigma$ correlation (see Section~\ref{sec:turb:perf:n-sigma}). Indeed, due to this correlation, $\eturb = \rho \sigma^2 / 2$ can be viewed as a passive marker of density, and therefore, will follow the dependence of $\rho$ on resolution. 

This behavior is particularly desirable for practical applications, where both $\rho$ and $\eturb$ often need to be used simultaneously. One example is the estimation of the local subgrid virial parameter, $\avir \propto \sigma^2/\rho \propto \eturb/\rho^2$ (see Equation~\ref{eq:avir} below), which can be used to predict the dynamical state of gas on subgrid scales and model local star formation efficiency in galaxy formation simulations (see Section~\ref{sec:galaxy}).
As long as the correlation between the density and $\eturb$ is captured by the subgrid model, it can be used to estimate such quantities despite the fact that the distributions of $\rho$ and $\eturb$ depend on the resolution.

To summarize, despite its simplicity, the SLES model captures many nontrivial features of small-scale turbulence in a direct simulation. It can correctly predict the global behavior of the mean turbulent energy, its dependence on scale, and the local correlations between the turbulent energy, gas density, and compression rate. In the next section, we investigate the model performance in a more practical application, a simulation of a disk galaxy.

\section{Galaxy Formation Simulations}
\label{sec:galaxy}

As an example of practical application, in this section, we explore the performance of the SLES model in a galaxy formation simulation. The model is used to dynamically predict the distribution of locally variable star-formation efficiency that depends on the properties of turbulence on unresolved scales. As we will show, the simple SLES model produces results close to those of a more complex explicit LES model.

\subsection{Galaxy Model and Methods}
\label{sec:galaxy:methods}

\begin{figure*}
\centering
\includegraphics[width=0.9\textwidth]{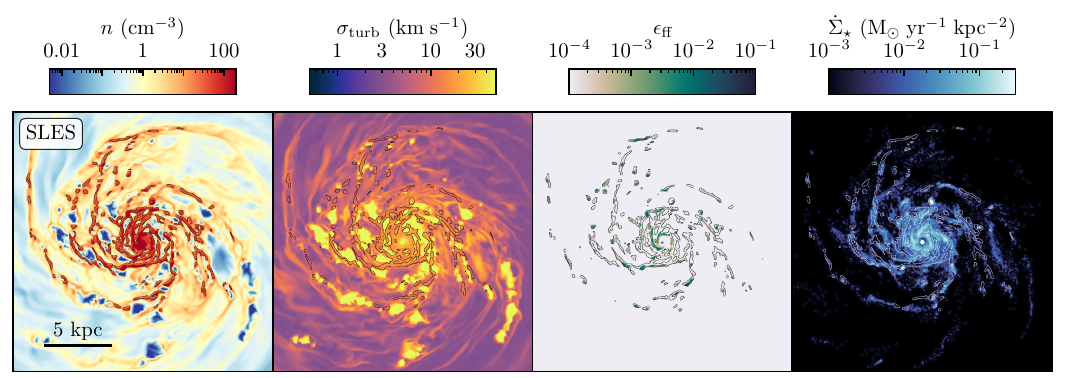}\\
\includegraphics[width=0.9\textwidth]{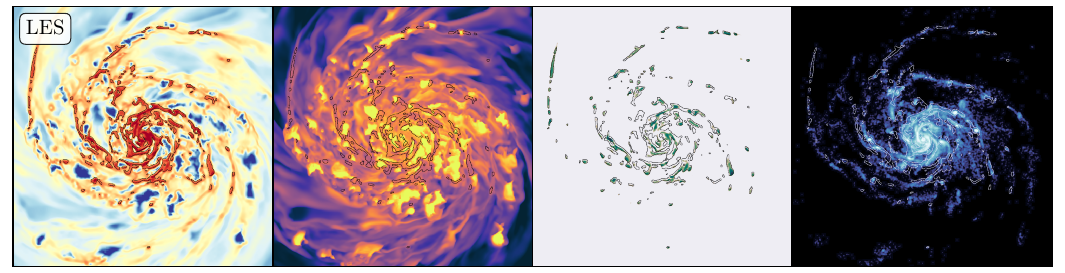}
\caption{\label{fig:galaxy-maps} Face-on view of the same galaxy resimulated with our new SLES (top row) and explicit LES (bottom panel) models for unresolved turbulence. Columns show, from left to right, slices of gas density ($n \equiv \rho/\mp$; for simplicity, we assume that the average molecular weight equals 1), subgrid turbulent velocity ($\st \equiv \sqrt{2\eturb/\rho}$), locally variable star formation efficiency per freefall time ($\epsff$; see Equation~\ref{eq:epsff-P12}), and the surface density of SFR computed using star particles younger than 100 Myr. To guide the eye, thin lines indicate density isocontours corresponding to $n = 20\cc$. Overall, both models produce similar distributions of $\st$, $\epsff$, and resulting $\SSFR$. The most noticeable differences are the higher levels of turbulence on the disk outskirts and in the center in the former, which originate partially due to the filtering of the bulk shear in the explicit model.}
\end{figure*}

We initialize an isolated $\Lstar$ disk galaxy simulation using one of the snapshots from the fiducial run of \citet{semenov17,semenov18,semenov19}, which was started from the initial conditions from the AGORA code comparison project \citep{agora-intro,agora-disk}. The computational mesh is adaptively refined by splitting the cells when their gas mass exceeds $\sim 8300\Msun$ or their local Jeans length (including the contributions from thermal and turbulent pressure) is resolved by fewer than 16 cells until the minimal cell size of $\Delta = 40\pc$ is reached. Gas cooling and heating are modeled using the \citet{gnedin12} model, assuming constant solar metallicity and a uniform radiation background corresponding to the H$_2$ photodissociation rate in the Lyman--Werner bands of $10^{-10}\;{\rm s^{-1}}$ \citep{stecher67}, which is suppressed in the dense gas following a shielding model calibrated against radiative transfer simulations of the ISM \citep[the ``L1a'' model from][]{safranekshrader17}. 
Stellar feedback is implemented by injecting density-, metallicity-, and resolution-dependent radial momentum and thermal energy calibrated by \citet{martizzi15}, with the momentum boosted by a factor of 5 to account for the effects of SN clustering \citep[e.g.,][]{gentry17,gentry19} and cosmic rays \citep{diesing18}, assuming the \citet{chabrier03} initial mass function.

In this paper, we rerun these simulations adopting different models for unresolved turbulence. The unresolved turbulence is dynamically tracked to predict the local dynamical state of gas and, coupled with it, locally variable star formation efficiency. Specifically, we use the common parameterization of the local star formation rate (SFR) density via the star formation efficiency per freefall time, 

\begin{equation}
\label{eq:rhoSFR}
\rhoSFR = \epsff \frac{\rho}{\tff}, 
\end{equation}
where $\tff = \sqrt{3\pi/32G\rho}$ and $\epsff$ depends exponentially on the local subgrid virial parameter following the model of \citet[][see \citealt{semenov16} for the explanation of the prefactor 0.9]{padoan12}:

\begin{equation}
\label{eq:epsff-P12}
\epsff = 0.9 \exp{(-\sqrt{\avir/0.53})},
\end{equation}
with the virial parameter for each simulation cell with size $\Delta$ defined as for a uniform sphere with radius $R = \Delta/2$ following \citet{bertoldi92}:

\begin{equation} 
\label{eq:avir}
    \avir \equiv \frac{5 \stot^2 R}{3GM} \approx 9.35 \frac{ (\stot/10\kms)^2 }{ (n/100\cc) (\Delta/40 \pc)^2}.
\end{equation}
The virial parameter accounts for both the thermal and subgrid turbulent pressure as $\stot \equiv \sqrt{ \cs^2 + \st^2 }$, with $\st \equiv \sqrt{2 \eturb/\rho}$ modeled using the subgrid turbulence model for $\eturb$.

In this paper, for comparison with the performance of the SLES model, we use our fiducial subgrid turbulence model with an explicit eddy-viscosity closure that we extensively tested in previous works \citep{semenov16,semenov17,semenov18,semenov19,semenov21-tf}, henceforth, the explicit LES model. The turbulence source term in this closure depends on the gradients of the \emph{fluctuating} component of the velocity field with the bulk flow velocity filtered out, the so-called ``shear-improved'' closure \citep{leveque07,schmidt14}. This is done to avoid spurious production of turbulence in constant shearing flows like differentially rotating disks.

In contrast to the shear-improved LES closure, the implicit source term in our SLES model provided by local numerical dissipation intrinsically depends on the gradients of the \emph{total} velocity field. Although the two models produce similar results (as presented below), the differences between the models can be partially explained by the shear being removed from our explicit LES model. Indeed, the resimulation of our LES run with the shear filtering turned off (i.e., with the closure relation dependent on the total velocity as in SLES) shows results intermediate between our semi-implicit and explicit models. Here we only present the comparison with the shear-improved model because this is the explicit model that is best suited for applications in galaxy simulations (see  Section~\ref{sec:discussion} for discussion).

\subsection{Distribution of Small-scale Turbulence}
\label{sec:galaxy:nsigma}

\begin{figure}
\centering
\includegraphics[width=\columnwidth]{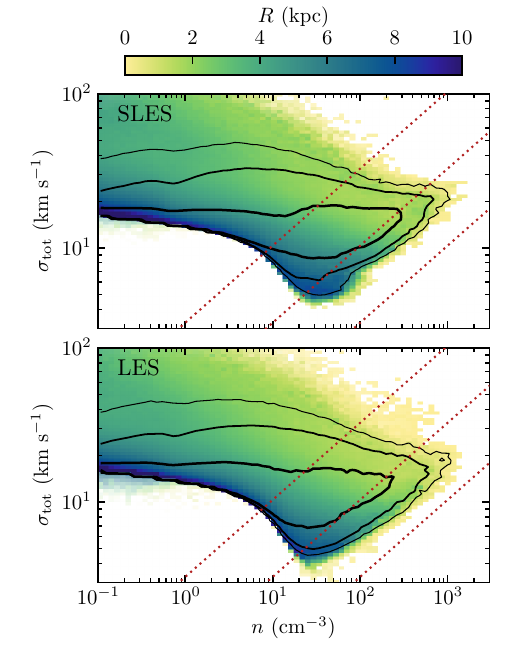}\\
\caption{\label{fig:galaxy-nsigma} Cell-by-cell distribution of the predicted total velocity dispersion, including subgrid turbulence and thermal motions, $\stot \equiv \sqrt{\st^2 + \cs^2}$, as a function of gas density, colored by the galactocentric radius of the cell. The panels correspond to the simulations with our new SLES (top) and explicit LES models (bottom). The diagonal dotted lines show the constant values of the local virial parameter: from top to bottom, $\avir = 100$, 10, and 1. The contours show 68\%, 95\%, and 99\% of cells weighted by mass. The data is stacked using 21 snapshots equally spaced between $t=500$ and $700\Myr$. The inner 500 pc, where the effects of shear are strong, is excluded from the plot. In the average ISM, both models produce similar distributions. }
\end{figure}

Figure~\ref{fig:galaxy-maps} shows the face-on view of our simulated galaxy with our new SLES (top row) and explicit LES (bottom panel) models for unresolved turbulence.  Qualitatively, both models produce similar global ISM structures. The two noticeable differences in the distribution of subgrid turbulent velocity (second column) are the higher level of turbulence on the outskirts and in the center with the SLES model compared to the explicit LES. In our resimulation of the LES run without shear filtering (not shown here), the level of turbulence in the outskirts and the center increases, indicating that this difference is, at least partially, produced by the shear due to differential rotation. Despite these differences, both models produce visually similar distributions of locally variable $\epsff$ (third column) and the resulting distribution of young stars ($<100\Myr$; rightmost column).

The distributions of star formation efficiencies and rates in the two models are similar because the local distributions of total velocity dispersion and gas density---which together define $\avir$ and $\epsff$---are similar. This is explicitly shown in Figure~\ref{fig:galaxy-nsigma}. In particular, the dotted lines in the plot indicate constant values of virial parameters, from top to bottom, $\avir = 100$, 10, and 1, corresponding to $\epsff \sim 10^{-8}$, 0.01, and 0.23, respectively.
Thus, it is particularly interesting that the distributions of $\stot$ are similar in the dense ($n > 10\cc$) gas because these are the densities at which most of the star formation occurs and where $\st > \cs \sim 1\kms$.

One of the physical reasons why these distributions are similar in dense regions is that they are largely shaped by the compressive heating (via the $PdV$ work) and decay of turbulence \citep{semenov16,semenov17}, and these processes are modeled in exactly the same way in both models (see Section~\ref{sec:model}). The evolution of gas in this part of the $n$--$\stot$ plane is, in fact, analogous to the evolution of gas in our supersonic developed turbulence tests in Figure~\ref{fig:n-sigma}: at high $\stot$, gas density on average increases, implying $\divg\vect{u} < 0$, while at low $\stot$, gas density decreases and $\divg\vect{u} > 0$, producing clockwise circulation of gas in that plane (see Figures 4 and 6 in \citealt{semenov17}).
Thus, as long as subgrid turbulence is produced at a reasonable level in diffuse gas, gas compression during the formation of dense regions will bring it to the values set by the interplay between the $PdV$ work and turbulence decay.

In diffuse gas at the disk outskirts, in contrast, the values of $\st$ are different (recall Figure~\ref{fig:galaxy-maps}). However, in these regions, turbulence is strongly subsonic, and $\stot$ is dominated by the thermal sound speed, which creates the lower envelope of the $n$--$\stot$ distribution at $n < 10\cc$. As a result, at these densities, small-scale turbulence does not affect gas dynamics, as its contribution to the total pressure is small. 

The inner 500 pc is excluded from the plot, as the strong shear in that region produces elevated $\st$ in the SLES model analogous to the artifact discussed in \citet[][see Figure 8 and related text in that paper]{semenov22}. The model behavior in the presence of strong shear, as in the galaxy center, should be treated with caution (see Section~\ref{sec:discussion}), in particular, due to the effects of stellar feedback associated with the differences in the SFR in the central region associated with such elevated $\st$. In addition, the gas behavior in the center of $\Lstar$ galaxies is expected to be strongly affected by active galactic nuclei, not modeled in this study. A detailed investigation of these effects requires a separate study.

\subsection{Global and Kiloparsec-scale Star Formation}

\begin{figure}
\centering
\includegraphics[width=\columnwidth]{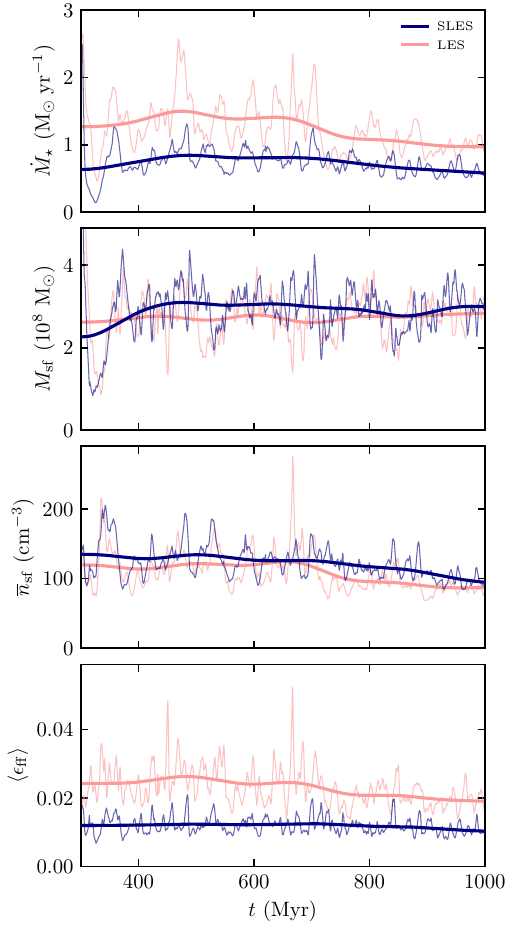}
\caption{\label{fig:galaxy-sfh} Evolution of global star formation properties in our galaxy simulations with two different models for unresolved turbulence. The panels show, from top to bottom, the total SFR, the total mass of star-forming gas, its average density, and the local star formation efficiency per freefall time, $\epsff$. Solid lines show the tracks smoothed on a 100 Myr timescale using a Gaussian filter. The SLES model results in $\sim 2$ times lower SFR than the explicit LES. This difference is mainly due to the difference in the predicted local $\epsff$, while the total mass and average density of star-forming gas are similar between the runs.}
\end{figure}

\begin{figure}
\centering
\includegraphics[width=\columnwidth]{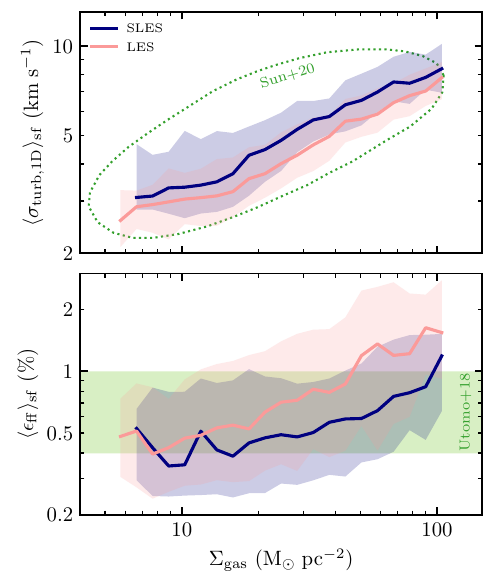}
\caption{\label{fig:galaxy-epsff} Predicted subgrid velocity dispersion (top) and star formation efficiency per freefall time (bottom) in star-forming gas within $1 \times 1 \kpc$ patches as a function of the total gas density. Only star-forming gas with $\epsff > 0.1\%$ is used in the averages; $\sigma$ is mass-weighted while $\epsff$ is SFR-weighted. Results of each simulation are stacked using 21 snapshots equally spaced between $t=500$ and $700\Myr$, with solid lines showing running medians and shaded regions showing 16$^\text{th}$--84$^\text{th}$ interpercentile ranges. For reference, the results from the PHANGS sample of nearby star-forming galaxies are shown with the dotted contour in the top panel \citep[][75\% of data points]{sun20} and with the green band in the bottom panel \citep{utomo18}. On average, the SLES model predicts higher $\st$ and lower $\epsff$ than the LES model, but the models lie within the scatter of each other and agree with the observed values.}
\end{figure}

Figure~\ref{fig:galaxy-sfh} shows the evolution of galaxy properties describing global star formation in our simulations: the global SFR ($\SFR$), the total mass of star-forming gas ($\Msf$), its average density ($\nsf$), and the SFR-weighted average local star-formation efficiency per freefall time ($\langle \epsff \rangle$). The differences in the global SFR can be explained via the differences in the other three quantities, as the SFR can be expressed as 
\begin{equation}
\label{eq:SFR}
    \SFR = \langle \epsff \rangle \frac{\Msf}{\overline{t}_{\rm ff}},
\end{equation}
where $\overline{t}_{\rm ff}$ is the inverse of $\langle \tff^{-1} \rangle$ weighted by $\epsff\,\rho\,dV$ for consistency with Equation~(\ref{eq:rhoSFR}), and $\nsf$ is the density corresponding to this freefall time.\footnote{This is one of the weighting schemes that make Equation~(\ref{eq:SFR}) consistent with (\ref{eq:rhoSFR}); for an alternative scheme, see, e.g., \citet{semenov24b}.}

With the SLES model, the global SFR is a factor of $\lesssim 2$ smaller than that in the simulations with the explicit LES model.
The lower three panels show that this effect is mainly due to the difference in the predicted average $\epsff$, while both the total amount and the average density of star-forming gas are approximately the same. The lower $\epsff$ in the SLES model can be partially explained by the higher level of turbulence generated by the shear in the differentially rotating disk.

Figure~\ref{fig:galaxy-epsff} compares the model predictions on spatially resolved kiloparsec scales. Specifically, the figure compares the scaling of the average subgrid velocity dispersion (top) and $\epsff$ (bottom) in star-forming gas as a function of the total gas density in $1\times 1$ kpc patches. The quantities on $y$-axes are averaged over the star-forming gas with $\epsff > 0.1\%$ with $\sigma$ being mass-weighted and $\epsff$ being SFR-weighted. 
In agreement with the above results, the SLES model predicts higher $\sigma$ and lower $\epsff$ in the star-forming gas than the explicit LES model. However, the distributions predicted in both models overlap significantly and are in the ballpark of the values observed in the PHANGS sample of galaxies from \citep{utomo17} and \cite{sun20}. 
The low $\epsff \sim 1\%$ values are a generic prediction of this $\avir$-dependent $\epsff$ model in both idealized isolated disk \citep{semenov16,semenov21-tf,polzin24} and cosmological simulations \citep{kretschmer20,segovia-otero24,semenov24b}, which results from the regulation of $\avir$ in star-forming gas to a narrow range due to stellar feedback \citep{polzin24}.
Note that the observed $\epsff$ range is provided in the figure only for reference, and the direct one-to-one comparison with simulation results requires a more careful analysis due to the evolutionary biases induced in the observational estimates of $\epsff$ \citep[e.g.,][]{segovia-otero24}.

The trend of $\sigma$ and $\epsff$ with $\Sgas$ is also interesting by itself. Both $\sigma$ and $\epsff$ increase with $\Sgas$. This result is nontrivial because, if $\sigma$ were to increase at fixed local gas density, the virial parameter, $\avir \propto \sigma^2/\rho$, would also increase, which would cause the decrease in $\epsff$ (see Equations~(\ref{eq:epsff-P12}) and (\ref{eq:avir})). Thus, the increasing trend of $\epsff$ with $\Sgas$ can be explained by the local gas density of star-forming gas also increasing with $\Sgas$, leading to the decrease of the virial parameter.

\section{Discussion}
\label{sec:discussion}

The new SLES model that we introduced combines features of both explicit and implicit LES. As an explicit LES, it carries the total turbulent energy on unresolved scales as an independent energy variable, enabling one to predict the local turbulent state of gas on unresolved scales. Unlike explicit LES, the source terms describing the turbulent cascade of energy from the resolved to subgrid scales are not modeled explicitly. Instead, the local numerical dissipation, calculated implicitly, is used as a source term.

This approach is motivated by the generic property of a turbulent flow that the dissipation rate is set by the rate of the kinetic energy cascade from large scales. In simulations where the physical dissipation scale is not resolved, the flow on the scales close to resolution is shaped by the numerical diffusivity of the adopted hydrodynamic method to ensure dissipation of the energy incoming from larger scales, making this diffusivity an effective model for the integral effect of the turbulent motions on unresolved scales. This idea constitutes the basis of the ILES.

Although at first glance the fact that numerical errors can conspire to produce a physical result might appear surprising, it has both physical and numerical foundations, which are extensively discussed in \citet{boris92,margolin02,grinstein}. In short, from the physical standpoint, the turbulent energy spectrum is sufficiently steep that the small-scale turbulence has only a weak dynamic effect on the resolved flow, alleviating resolution requirements for modeling the largest scales of turbulent flows. At the same time, the spectrum is not too steep, so that the small-scale motions can efficiently mix inhomogeneities produced on larger scales. Except for strong shocks (which we discuss below), the kinetic energy also cascades through all inertial-range scales, making the model applicable as long as the onset of the turbulent cascade is resolved. To satisfy these properties, the numerical method is required to respect the generic properties of the underlying hydrodynamic equations---such as conservation, causality, and lack of spurious oscillations---to ensure that the kinetic energy cascades at the correct rate from large to small scales, where numerical diffusivity dissipates this energy at the same rate. 

It is worth noting that both explicit LES and ILES rely on the idea of a local kinetic energy cascade continuous in spatial scales. The notion of cascade becomes nuanced for supersonic turbulence due to compressibility effects. In general, cascades in both kinetic and thermal energies need to be considered in this case as they can exchange energy on any scale as a result of the $PdV$ work done by the gas during compression and expansion (pressure dilatation; see, e.g., \citealt{aluie13,eyink18}; see also \citealt{garnier} for the review of LES models for compressible turbulence). One special case of compressible turbulence for which the presence of a kinetic energy cascade has been derived analytically is the isothermal case \citep{aluie12}. For this reason, the above issue is also alleviated in ISM and galaxy formation applications, as most of the volume is typically occupied by the subsonic or transonic, weakly compressible hot ($T > 10^5$ K) and warm ($T \sim 10^4$ K) media, while most of the mass and star formation resides in the cold ($T \lesssim 100$ K) medium, which, albeit supersonic, is nearly isothermal due to strong radiative losses.

The key difference between our semi-implicit method and ILES is that the latter does not explicitly distinguish between subgrid turbulent and thermal energies. Owing to energy conservation, in ILES, all dissipated energy instantly appears as heat, bypassing the unresolved turbulent cascade. In contrast, in SLES, we track subgrid turbulent and thermal energies separately and explicitly model the decay of turbulence into heat. For adiabatic flows, SLES and ILES produce identical results as both $\eturb$ and $\eth$ are described with an ideal gas equation of state with $\gamma = 5/3$, making the total pressure independent of the partitioning between $\eturb$ and $\eth$. In the presence of cooling and heating, however, the results are different because only $\eth$ is subject to these processes, while $\eturb$ dissipates on a different timescale, close to cell-crossing (see Section~\ref{sec:model:decay}). For example, in dense ISM, $\eth$ is subject to strong radiative losses, while $\eturb$ can retain some of the dissipated energy, delaying cooling and providing local nonthermal support in the cold ISM. Note that, unlike turbulent viscosity, this turbulent pressure must be explicitly included in hydrodynamic equations, analogously with thermal pressure.

The small-scale (close-to-resolution) structure of the flow and the numerical dissipation scale where the turbulent power spectrum truncates strongly depend on the resolution and the details of the numerical solver \citep[e.g.,][]{kitsionas09,kritsuk11}. This implies that the effective spatial scale at which the subgrid turbulent energy is estimated in our method---set by the numerical dissipation scale---will vary with the details of the hydrodynamic method, implying that the turbulence decay rate needs to be recalibrated when used in different codes (see Section~\ref{sec:turb:calib}). In the context of ILES and SLES, the second-order finite-volume methods are particularly attractive because the leading (third) order of their truncation errors matches that of the explicit subgrid terms in LES models \citep[][]{margolin02}. The ART code used in this paper is one example of such a second-order finite-volume hydrodynamic code.

In such second-order accurate methods, numerical dissipation starts acting on relatively large scales, $\sim$30 cells, and dominates on $\sim$5--10 cells \citep[e.g.,][]{kitsionas09,federrath10,federrath11,kritsuk11}. 
\newtext{For example, \citet{grete23} quantified the resolution dependence of the numerical dissipation scale using energy transfer analysis and reported that dissipation peaks at $\sim$5--10 cells, with a tail extending to scales up to $\sim$5 times larger (see their Figure 4 and Table 1). In the context of the current study, these results imply that the subgrid turbulent energy predicted by the SLES model should be interpreted as the kinetic energy on scales smaller than the dissipation scale, convolved with the code-specific scale dependence of dissipation.}

The flow and the distribution of $\eturb$ become smoothed by the numerical diffusivity on a similar scale (recall Figure~\ref{fig:turbbox-maps}). However, the resulting distribution of $\eturb$ can still be used to predict the distributions of turbulence properties, such as $\avir$ or Mach numbers. This is because the subgrid model captures the correlations between $\eturb$ and other hydrodynamic properties (recall Figure~\ref{fig:n-sigma}), even though they are all affected by numerical diffusivity. 
The predicted distributions of $\avir$ or Mach numbers, in turn, can be used to improve the subgrid modeling of processes dependent on the turbulent structure of gas. In the context of galaxy formation, these processes include star formation (see Section~\ref{sec:galaxy}; see also \citealt{braun15,semenov16,semenov21-tf,kretschmer20}); turbulent diffusion of metals \citep[e.g.,][]{rennehan21}; cooling, heating, and chemical reaction rates; acceleration and transport of cosmic rays; and microscopic dynamo \citep[e.g.,][]{liu22}, among others.

The existing studies of the effects of unresolved turbulence listed above rely on explicit LES models.
The key advantage of SLES over explicit models is the simplicity of implementation, as it reuses the functionality present in many hydrodynamic codes used for galaxy formation, specifically the dual-energy formalism \citep[see Section~\ref{sec:model:iles};][]{ryu93,bryan95}. In this formalism, thermal energy is modeled twice: as a part of conserved total energy and separately, following an adiabatic equation, enabling one to compute gas temperature in strongly supersonic regions where the difference $\etot - \ekin = \eth \ll \etot$ becomes numerically unreliable. SLES can be implemented by introducing an additional energy variable describing unresolved turbulence and setting it to $\eturb = \etot - \ekin - \eth$ after each step. As $\etot$ is conserved and $\eth$ is modeled adiabatically, this step implicitly puts all dissipated energy into $\eturb$ \citep[see also][]{semenov22}. 

\newtext{As a result, the SLES model imposes negligible costs in terms of both computation and memory. The computational overhead is minimal because the energy cascade term reuses calculations already performed in the code without an LES model (Section~\ref{sec:model:sles}), while all other source and sink terms are either described by analytical functions (e.g., turbulence decay and SN driving) or treated in exactly the same way as in the explicit LES model (e.g., turbulent diffusion; Section~\ref{sec:model:decay}). The memory requirements are also low, since only one additional state variable is needed to track $\eturb$ in each cell.} In principle, one may \newtext{not introduce a new variable for $\eturb$ at all} and instead apply all appropriate terms to $\eth$ and use the above equation to calculate $\eturb$ when needed. However, we choose to follow $\eturb$ as a separate variable as it enables us to compute it in regions where the above difference is dominated by numerical errors, by analogy with the original dual-energy formalism.

The SLES model can \newtext{potentially} be generalized in the presence of magnetic fields. In this case, $\eturb$ should be interpreted as the subgrid kinetic plus magnetic energy, while the implicit sourcing by numerical dissipation would encompass the integral effect of LES-like viscosity and resistivity terms entering the MHD equations \citep[see, e.g.,][for an example of such explicit subgrid MHD models]{grete16,grete17,vlaykov16}. 
\newtext{Note, however, that the validity of such models must be assessed with care as some of the basic assumptions of the LES framework---the scale locality and unidirectionality of the energy cascade---can be violated in the presence of magnetic fields \citep[e.g.,][]{alexakis07,yousef07,stepanov15,grete17-mhdcascade,schekochihin22,friedrich24,beattie25}. First, magnetic fields can enable nonlocal energy transport, breaking the assumption that local source terms for subgrid magnetized turbulence can depend solely on the local flow properties. Second, they enable an inverse cascade, in which energy flows from small, unresolved to large, resolved scales. In the context of the SLES model, the inverse cascade acts analogously to locally negative dissipation: the entropy of the resolved flow decreases as kinetic energy is transferred from unresolved to larger scales. Such negative dissipation is present in most second- and higher-order methods, but it is purely numerical in origin and must be treated with caution (see the discussion of entropy stability below). \newtextb{An MHD extension of the SLES model would therefore require an explicit treatment of the energy transfer behavior of the magnetic fields across the resolution scale, which can be particularly challenging for low-order methods, since such a model would need a large spatial kernel ($>10$ cells) to fully capture the dissipative scales where this transfer occurs numerically \citep[see, e.g.,][]{grete17}.}}

The results of Section~\ref{sec:galaxy} show that the distribution of $\eturb$ in the cold, supersonic ISM is only weakly sensitive to the choice of model. This is because in this regime the production of small-scale turbulence is dominated by the heating of preexisting turbulence via the $PdV$ work exerted by gas undergoing compression in spiral arms, collision of SN-driven shells, and disk instabilities \citep[see][]{semenov16,semenov17}. Thus, the choice of the model for the cascade term, $\Sigma$, in Equation~(\ref{eq:eturb}) might not be as crucial for predicting the turbulent properties of the cold ISM, as long as the adiabatic $PdV$ term and turbulence decay, $\epsilon$, are modeled appropriately. However, significant differences do arise in regions where the cascade of energy from resolved scales dominates the sourcing of $\eturb$. 

One example is the disk outskirts and very center, where the SLES model predicts higher turbulent velocities than the explicit LES model (Section~\ref{sec:galaxy}). One reason for this discrepancy is the difference in the treatment of shearing flows. Our explicit model adopts a shear-improved LES closure where the bulk flow is filtered out and only the fluctuating part of the velocity field is interpreted as a part of the turbulent cascade \citep[][see also Section~\ref{sec:model:les}]{leveque07}. To avoid spurious production of $\eturb$ in stable shearing flows, such as differentially rotating disks, one needs to adopt a similar treatment in the SLES model. For example, instead of putting all dissipated energy into $\eturb$, one can make it a function of the difference between the local filtered and total flows, such that in purely shearing flows, kinetic energy is dissipated directly into heat. We leave the investigation of such models to future work.

Another potential reason why the SLES and explicit LES models produce different results in the subsonic regime of the disk outskirts is numerical. In this regime, $\eturb$ becomes relatively small compared to other energy terms in Equation~(\ref{eq:eturb-implicit-source}), making the estimate numerically unreliable. In fact, the positivity of the implicit source term in that equation is not generally guaranteed. To guarantee this positivity, the hydrodynamic method must be entropy-stable, i.e., the local entropy must not decrease, or there must be no negative dissipation \citep[e.g.,][]{tadmor03,schmidtmann17}. In our turbulent box tests, we find that such ``negative dissipation'' regions occur in the low-density gas next to strong density gradients, but they constitute a negligible fraction of the total volume and energy budget. Most of the dissipation is positive, and it occurs in high-density converging flows, leading to a strong positive correlation of $\eturb$ with gas density (recall Figure~\ref{fig:turbbox-maps}). These effects, however, can become important for modeling subgrid turbulence in subsonic, weakly compressible regimes, which we leave to a separate study.

Apart from shearing flows, another example where the dissipated energy can bypass a turbulent cascade and be thermalized directly is strong shocks. To this end, one can use a shock-finding algorithm and make the fraction of thermalized energy dependent on the presence and local properties of the shock. In our current implementation of the SLES model, this is not taken into account, and strong shocks convert kinetic energy into $\eturb$ instead. Note, however, that this does not affect the shock solution because the total postshock pressure remains identically the same, independently of the partitioning between $\eturb$ and $\eth$ as both are described by the ideal gas equation of state with the same adiabatic index of $5/3$. In addition, our tests in the decaying supersonic turbulence setup show that our simpler model works reasonably well (Section~\ref{sec:turb}).

\section{Summary}
\label{sec:summary}

We have introduced a dynamic model for unresolved turbulence in hydrodynamic simulations that explicitly tracks subgrid turbulent energy by sourcing it implicitly via numerically dissipated kinetic energy in each cell. This model is physically motivated by a generic property of turbulent flows that they dissipate kinetic energy at a rate set by the energy cascade rate from large scales, which is independent of fluid viscosity, regardless of its nature, be it physical or numerical. For this reason, the local numerical dissipation rate can be used as a reasonable approximation for the local rate of energy cascade from resolved to subgrid scales.

The model combines features of both explicit Large Eddy Simulations (LES; Section~\ref{sec:model:les})---as it explicitly tracks the evolution of subgrid turbulence and its decay into heat---and implicit LES (ILES; Section~\ref{sec:model:iles}), where unresolved scales are represented by numerical dissipation of the hydrodynamic method. To highlight these analogies, we dubbed our model Semi-implicit LES (SLES; Section~\ref{sec:model:sles}).

In its simplest form, presented here, this model involves only one tunable parameter that describes subgrid turbulence decay into heat. We calibrate this parameter against high-resolution simulations of decaying supersonic turbulence (Section~\ref{sec:turb:calib}) and then test its predictions in a turbulent box (Section~\ref{sec:turb:perf}) and isolated galaxy disk simulations (Section~\ref{sec:galaxy}). Despite its simplicity, the model can capture many nontrivial properties of these turbulent flows. These results are summarized as follows:

\begin{enumerate}
    \item In the decaying supersonic turbulence test, the model predicts the correct evolution of average small-scale turbulent energy and its dependence on the spatial scale (Figure~\ref{fig:var-scale}). 

    \item The model correctly predicts both the slope and scatter of the correlation between small-scale turbulence and gas density (Figure~\ref{fig:n-sigma}; see also Figure~\ref{fig:turbbox-maps} for visual impression).

    \item The model captures the dependence of small-scale turbulence on the local compression rate at fixed density: more turbulent regions correspond to regions with stronger compression (Figures~\ref{fig:n-sigma} and \ref{fig:nsigma-scatter}). Interestingly, this correlation is analogous to the behavior of supersonic dense ISM in galaxy simulations (see Section~\ref{sec:galaxy:nsigma}).

    \item Small-scale turbulence exhibits two limiting regimes of density dependence. In weak turbulence, cell-crossing time is long, and therefore turbulence decays slowly and behaves near-adiabatically ($\eturb \propto \rho^{5/3}$, or $\st \propto \rho^{1/3}$). In strong turbulence, cell-crossing time is short, and turbulence behaves near-isothermally ($\eturb \propto \rho$, or $\st = {\rm const}$). The $\st$--$\rho$ relation slope in our decaying supersonic turbulence test exhibits intermediate values, indicating that the small-scale turbulence is, on average, in an intermediate regime (Figure~\ref{fig:n-sigma}).

    \item In galaxy simulations, the new model produces the distribution of $\st$ similar to a more sophisticated explicit LES model, especially in the cold supersonic ISM responsible for star formation (Figures~\ref{fig:galaxy-maps} and \ref{fig:galaxy-epsff}). On the disk outskirts and in the center, however, the model produces a higher level of turbulence, indicating that modeling of strongly shearing and subsonic flows with this model requires further investigation (see Section~\ref{sec:discussion})

    \item Despite these differences, the effect on the predicted star formation rates is only a factor of $\sim$2, which is within the uncertainty of existing star formation and feedback models (Figure~\ref{fig:galaxy-sfh}). Both models agree in terms of predicted small-scale turbulent velocities and star formation efficiencies in star-forming regions, which are also in the ballpark of the observed values (Figure~\ref{fig:galaxy-epsff}).
\end{enumerate}

Our results indicate that the new SLES model can produce realistic results---at least in the context of modeling supersonic star-forming ISM---at a fraction of the cost of explicit LES. The model is significantly more straightforward to implement in existing codes used to carry out ILES as it largely relies on the already implemented infrastructure. This warrants further investigation of the model performance in other regimes (such as shearing and subsonic flows pointed out above) and different hydrodynamic codes.

\section*{Acknowledgements}
I would like to thank the anonymous referee for constructive comments, which helped improve the manuscript.
I am also grateful to Christoph Federrath, Troels Haugb{\o}lle, Lars Hernquist, Ralf Klessen, Andrey Kravtsov, Alexei Kritsuk, Jonathan Stern, Romain Teyssier, and the participants of the Ringberg workshop ``Advancements in Computational Galaxy Formation'' for insightful discussions.
I also thank Alexei Kritsuk for providing access to the initial conditions of an isothermal turbulence test used in the code comparison project of \citet{kritsuk11}.
The analyses and simulations presented in this paper were performed on the NASA Pleiades cluster supported by the NASA High End Computing (HEC) Program through the NASA Advanced Supercomputing (NAS) Division at Ames Research Center and on the FASRC Cannon cluster supported by the FAS Division of Science Research Computing Group at Harvard University.
Support for V.S. was provided by NASA through the NASA Hubble Fellowship grant HST-HF2-51445.001-A awarded by the Space Telescope Science Institute, which is operated by the Association of Universities for Research in Astronomy, Inc., for NASA, under contract NAS5-26555, and by Harvard University through the Institute for Theory and Computation Fellowship.
The analyses presented in this paper were greatly aided by the following free software packages: {\tt NumPy} \citep{numpy_ndarray}, {\tt SciPy} \citep{scipy}, {\tt Matplotlib} \citep{matplotlib}, and {\tt yt} \citep[][]{yt}. We have also used the Astrophysics Data Service (\href{http://adsabs.harvard.edu/abstract_service.html}{\tt ADS}) and \href{https://arxiv.org}{\tt arXiv} preprint repository extensively during this project and the writing of the paper.

\appendix

\section{Dependence on Resolution}
\label{app:dns-convergence}

\begin{figure}
\centering
\includegraphics[width=\columnwidth]{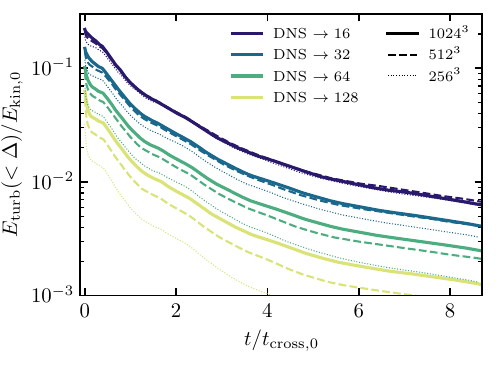}
\caption{\label{fig:dns-res} Convergence of the average $\eturb$ estimate from DNS with resolution (line styles) and coarse-graining scale (line colors). Our fiducial $1024^3$ DNS is converged for $64^3$ and all coarser grids, while at $128^3$ significant deviations become noticeable as it probes $\eturb$ on the scales affected by numerical dissipation: $1024/128 = 8$ cells.}
\end{figure}

The turbulent cascade in simulations without explicit viscosity truncates at the numerical dissipation scale close to the resolution. As Figure~\ref{fig:power-spectra} shows, in our high-resolution ($1024^3$) DNS that we use to calibrate and test our SLES model, this truncation starts around $k/k_{\rm min} = L_{\rm box}/L \sim 100$, which corresponds to $\sim$10 cells, in agreement with the \citet{kritsuk11} results. Only the estimates of small-scale turbulence, $\eturb$, on sufficiently larger scales should be used for comparisons with SLES, as $\eturb$ on smaller scales is not numerically converged. 

In Figure~\ref{fig:dns-res}, we investigate the appropriate coarse-graining scales that can be used for our purposes. Different line styles show DNSs with different resolutions, while the line colors show the average $E_{\rm turb} \equiv \langle \eturb \rangle_{\rm box}$ coarse-grained using different coarse grids: $16^3$, $32^3$, $64^3$, and $128^3$ (see Section~\ref{sec:turb:setup}). A finer coarse grid corresponds to the estimate of $\eturb$ over a smaller number of mesh cells, e.g., for our fiducial $1024^3$ DNS, these grids correspond to 64, 32, 16, and 8 cells per side. 

The comparison of DNSs with different resolutions and spacing between results for different coarse grids shows that the results of our DNS run are converged when considered on down to $64^3$ coarse grids. In Section~\ref{sec:turb}, however, we also show the results for $128^3$, which should be interpreted with caution as they probe $\eturb$ on scales close to the numerical dissipation scale.

\bibliographystyle{aasjournal}
\bibliography{}

\end{document}

%% file: citation_fix.tex
\usepackage{etoolbox}
\makeatletter
\patchcmd{\NAT@citex}
  {\@citea\NAT@hyper@{\NAT@nmfmt{\NAT@nm}\NAT@date}}
  {\@citea\NAT@nmfmt{\NAT@nm}\NAT@hyper@{\NAT@date}}
  {}
  {}
\patchcmd{\NAT@citex}
  {\@citea\NAT@hyper@{%
     \NAT@nmfmt{\NAT@nm}%
     \hyper@natlinkbreak{\NAT@aysep\NAT@spacechar}{\@citeb\@extra@b@citeb}%
     \NAT@date}}
  {\@citea\NAT@nmfmt{\NAT@nm}%
   \NAT@aysep\NAT@spacechar%
   \NAT@hyper@{\NAT@date}}
  {}
  {}
\patchcmd{\NAT@citex}
  {\@citea\NAT@hyper@{%
     \NAT@nmfmt{\NAT@nm}%
     \hyper@natlinkbreak{\NAT@spacechar\NAT@@open\if*#1*\else#1\NAT@spacechar\fi}%
       {\@citeb\@extra@b@citeb}%
     \NAT@date}}
  {\@citea\NAT@nmfmt{\NAT@nm}%
   \NAT@spacechar\NAT@@open\if*#1*\else#1\NAT@spacechar\fi%
   \NAT@hyper@{\NAT@date}}
  {}
  {}
\makeatother

%% file: defs.tex
\newcommand{\vect}[1]{{\pmb #1}}
\newcommand{\dotp}{\boldsymbol{\cdot}}
\newcommand{\grad}{\boldsymbol{\nabla}}

\newcommand{\divg}{\grad \dotp}

\def\eth{e_{\rm th}}

\def\Pturb{P_{\rm turb}}
\def\eturb{e_{\rm turb}}

\def\ekin{e_{\rm kin}}
\def\Eturb{E_{\rm turb}}

\def\etot{e_{\rm tot}}

\def\stot{\sigma_{\rm tot}}
\def\st{\sigma_{\rm turb}}
\def\cs{c_{\rm s}}
\def\Mach{\mathcal{M}}
\def\mp{m_{\rm p}}

\def\tcross{t_{\rm cross,0}}

\def\tff{t_{\rm ff}}

\def\Ekininit{E_{\rm kin,0}}
\def\Lbox{L_{\rm box}}
\def\Ngrid{N_{\rm grid}}

\def\epsff{\epsilon_{\rm ff}}
\def\avir{\alpha_{\rm vir}}

\def\Sgas{\Sigma_{\rm gas}}
\def\SSFR{\dot\Sigma_{\rm \star}}
\def\rhoSFR{\dot\rho_{\rm \star}}

\def\SFR{\dot{M}_{\rm \star}}
\def\Msf{M_{\rm sf}}
\def\nsf{\overline{n}_{\rm sf}}

\def\ce{C_\epsilon}
\def\ceprime{C^\prime_\epsilon}

\def\Lstar{L_\star}

\def\cc{\;{\rm cm^{-3}}}
\def\kms{\;{\rm km\;s^{-1}}}
\def\Msun{\;M_\odot}
\def\pc{\;{\rm pc}}
\def\kpc{\;{\rm kpc}}
\def\Myr{\;{\rm Myr}}